\documentclass[10pt]{elsarticle}
\usepackage{xcolor}
\definecolor{darkyellow}{RGB}{200,200,0}

\usepackage [autostyle, english = american]{csquotes}

\usepackage[english]{babel} 

\usepackage{graphicx} %
\usepackage{amssymb}
\usepackage{amsmath}
\usepackage{caption}
\usepackage{subcaption}
\usepackage{bm}
\usepackage{algorithm}
\usepackage{algpseudocode}
\usepackage{listings}
\usepackage{todonotes}
\usepackage[normalem]{ulem}

\newcommand{\mydel}[1]{}
\usepackage[pdfcreator = {pdflatex},
            hyperindex = {true},
            colorlinks = {true},
            linkcolor  = {blue},
            citecolor  = {blue},
            urlcolor   = {blue}]{hyperref}
\usepackage{cleveref}
\usepackage{multirow}
\usepackage{booktabs}
\usepackage{pgfplots}
\pgfplotsset{compat=1.18}
\usepackage{mathtools}
\usepackage[section]{placeins}
\usepackage{type1cm}

\allowdisplaybreaks

\newcommand{\x}{\bm{x}}
\newcommand{\y}{\bm{y}}

\newcommand{\ux}{\bm{u}}

\usetikzlibrary{arrows.meta}
\usetikzlibrary{patterns}
\usetikzlibrary{calc}

\begin{document}
\begin{frontmatter}
\title{A scalable Ewald-free BIE framework for periodic Stokes flow via hierarchical proxy sums}
 \author[1]{Tianyue Li}
\author[2,*]{Dhairya Malhotra}
\author[1]{Shravan Veerapaneni}
\affiliation[1]{Department of Mathematics, University of Michigan}
\affiliation[2]{Flatiron Institute, Simons Foundation}
\affiliation[*]{Corresponding author; dmalhotra@flatironinstitute.org}
\date{}
\begin{abstract}
Particulate Stokes flow in confined, periodic geometries underlies a broad class of problems in biophysics, microfluidics, and the rheology of complex fluids. Boundary integral equation (BIE) methods are a natural tool for such problems, but existing periodization schemes rely either on periodic Green's functions, which are restrictive for complex confining geometries, or on free-space schemes that solve auxiliary proxy strengths alongside the surface densities in an extended linear system whose cost scales unfavorably in three dimensions. We present a BIE framework for three-dimensional particulate Stokes flow in periodic pipes with circular cross-sections, wall-bounded doubly-periodic, and triply-periodic geometries that uses only the free-space Green's function and avoids both Ewald summation and the extended linear system. Proxy sources placed on equivalent surfaces of the kernel-independent FMM (KIFMM) form the auxiliary basis, and contributions from far image boxes are captured by a hierarchical proxy sum made absolutely convergent by a net-force-zero compatibility condition. The resulting periodization precomputation depends only on the periodic-box geometry, independent of the kernel and of the surfaces inside the box, and is reused verbatim across the Stokeslet, stresslet, and rotlet. Combined with high-order adaptive surface discretizations, the method achieves high-order accuracy at $\mathcal{O}(N)$ cost with a single layer of image boxes in the near field. Numerical examples on dense polydisperse suspensions with thousands of particles and on flow through complex periodic channels, together with strong and weak scaling studies, demonstrate efficient performance on systems with millions of degrees of freedom on distributed-memory architectures.
\end{abstract}
\end{frontmatter}

\section{Introduction}
 \label{sec:intro}
 Particulate Stokes flow---the inertia-free motion of rigid and deformable bodies through a viscous fluid---arises across a broad spectrum of natural and engineered systems, from cellular biomechanics and microorganism motility to lab-on-a-chip processing and the rheology of dense suspensions. Representative examples include polydisperse colloidal suspensions and sediments \cite{maxey2017simulation,ness2022physics}; the microcirculation of red blood cells, vesicles, and other deformable bodies through vascular and lymphatic networks \cite{freund2014numerical, secomb2017blood}; microfluidic devices for high-throughput cell sorting, deformability cytometry, and the separation of circulating tumor cells \cite{jin2014technologies, darling2015highthroughput}; and suspensions of self-propelled particles and motor-driven cytoplasmic flows \cite{yan2020scalable,alert2022active, nazockdast2017cytoplasmic}. Such applications demand simulations that resolve fine geometric features---a microswimmer's prescribed-slip surface, a vesicle membrane, the slender gap between nearly touching colloids---while capturing the long-range, many-body hydrodynamic interactions of large particle ensembles over time horizons sufficient to reach a steady state or to extract reliable ensemble averages. Small errors in simulating the particle dynamics accumulate over these horizons and bias macroscopic observables, including basic quantities such as the osmotic pressure of a dense suspension \cite{foss2000brownian}. Periodic boundary conditions are the standard idealization in this regime: a triply-periodic unit cell is the canonical representative volume for bulk rheology and homogenization, while doubly- and singly-periodic configurations with confining walls model microfluidic channels and vascular segments \cite{marple2016fast, barnett2018unified}; even when the underlying system is not strictly periodic, this idealization remains the standard route to simulating a representative sample at tractable cost. This paper addresses the corresponding computational task: simulation of transient particulate Stokes flow in three-dimensional periodic geometries to high accuracy and at scalable cost.

Boundary integral equation (BIE) methods provide a natural framework for these problems: they recast the governing PDE as an integral equation on the bounding surfaces \cite{youngren1975stokes, pozrikidis1992boundary}, with the resulting dense matrix--vector products computable in $\mathcal{O}(N)$ work using fast algorithms such as the Fast Multipole Method (FMM) \cite{Greengard1987}. To impose periodicity, the standard approach is to substitute the free-space kernel with its periodic counterpart, written as a lattice sum over translates of the source. The single-layer potential on a surface $\Gamma$ then takes the form
\begin{equation}
\boldsymbol{u}(\boldsymbol{x}) \,=\, \sum_{\boldsymbol{p}\in\mathbb{Z}^d}\,\int_{\Gamma} \boldsymbol{S}\big(\boldsymbol{x},\boldsymbol{y}+g(\boldsymbol{p})\big)\,\bm{\mu}(\boldsymbol{y})\,d\Gamma(\boldsymbol{y}), \label{periodicSum}
\end{equation}
where $d$ is the dimension of periodicity, $g$ maps lattice index $\boldsymbol{p}$ to the corresponding periodic image (e.g., $g(\boldsymbol{p})=L\boldsymbol{p}$ for triply-periodic systems with period $L$), $\boldsymbol{S}$ is the free-space Stokeslet, and $\bm{\mu}$ is the density on $\Gamma$ \cite{lindbo2012fast}.

\begin{figure}[!t]
    \centering
    \includegraphics[width=0.9\linewidth]{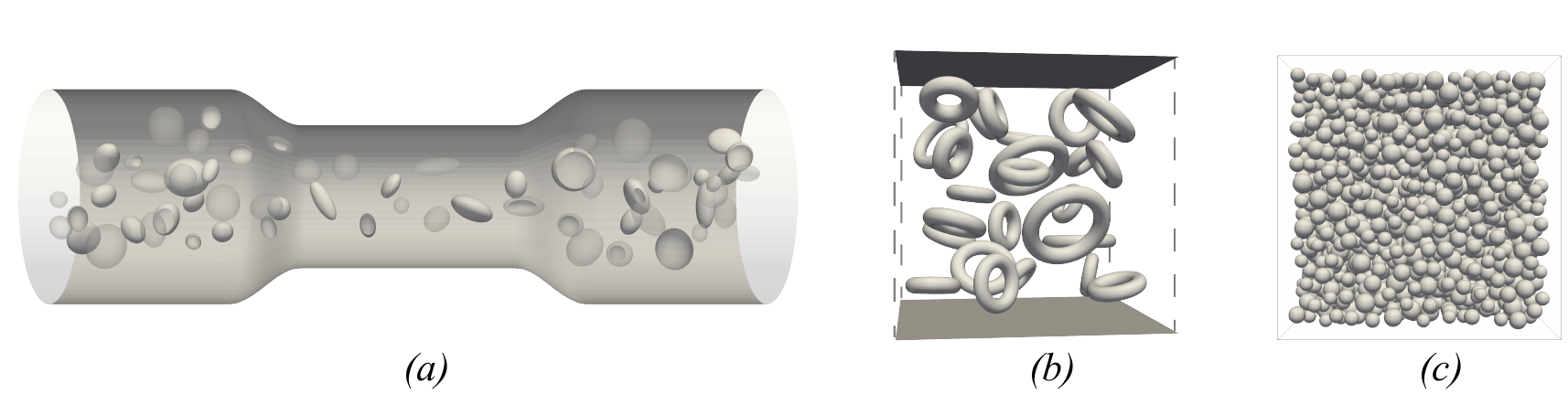}
    \caption{\small Three representative periodic geometries considered in this work. (a) Singly-periodic: a converging-diverging channel packed with polydisperse slip-driven spheroidal particles. (b) Doubly-periodic: passive toroidal particles in pressure-driven flow between two parallel no-slip walls. (c) Triply-periodic: flow over a dense polydisperse spherical particle suspension; one evaluation of the BIE operator for this unit cell takes under 1.5 seconds on 1920 CPU cores.}
    \label{fig:all_periodicity_examples}
\end{figure}

Direct BIE realizations of \eqref{periodicSum} for Stokes have been developed in \cite{Zick1982Stokes, pozrikidis1996computation, fan1998completed, greengard2004integral}, while FFT-based Ewald methods \cite{hasimoto1959periodic, saintillan2005smooth, lindbo2010spectrally, af2014fast, wang2016spectral} accelerate the underlying $N$-body sum through an analytic short- and long-range splitting whose smooth long-range part is evaluated by FFT. These methods remain widely used, but they face well-known difficulties: closed-form periodic Green's functions are rarely available for arbitrary confining geometries; uniform Fourier grids preclude spatial adaptivity for multiscale or polydisperse distributions; and performance deteriorates on high-aspect-ratio period boxes, where the lattice sum in \eqref{periodicSum} converges only on a ball, forcing many additional near-field images.

Recent work has narrowed these gaps. Spectral Ewald has been extended to free space \cite{afklinteberg2017free} and to arbitrary periodicity, including wall-bounded singly- and doubly-periodic Stokes problems \cite{bagge2023arbitrary}. The dual-space multilevel kernel-splitting (DMK) framework \cite{jiang2025dmk,afklinteberg2025dmk} replaces the FFT in the far field with a telescoping multilevel splitting that is natively adaptive on a tree, largely lifting the spatial-adaptivity limitation for kernels admitting such a splitting. A prolate-spheroidal-wave-function splitting of the biharmonic kernel \cite{afklinteberg2025dmk} roughly halves the Fourier bandlimit required at high precision, accelerating both Spectral Ewald and DMK for the Stokeslet, stresslet, and elastic kernels. Two structural features of the Ewald-class pipeline nevertheless persist and motivate the present approach: each new kernel requires its own splitting derivation, and partially periodic, wall-bounded geometries with curved walls are not natively handled, since Fourier-domain splittings are most natural in unbounded boxes.

A complementary line of work avoids the periodic Green's function entirely, using only the free-space kernel and captures far-image contributions through a small auxiliary basis of proxy sources outside the unit cell, with periodicity imposed as a linear constraint. The idea originated for two-dimensional quasi-periodic Helmholtz scattering \cite{barnett2011new} and was developed further through fast direct solvers and a three-dimensional acoustic-scattering analog for axisymmetric obstacles \cite{gillman2013fast, cho2015robust, liu2016efficient}. Marple et al.~\cite{marple2016fast} adapted it to BIE periodization for two-dimensional Stokes flow of vesicle suspensions in arbitrary-shape channels, and Barnett et al.~\cite{barnett2018unified} generalized this to a unified scheme for {\em doubly}-periodic Laplace and Stokes BVPs in 2D, accommodating many inclusions per unit cell, skew unit cells, adaptivity, and close-to-touching geometries via spectral quadratures.

In all of these works, proxy strengths are solved alongside the wall layer densities in a single \emph{extended linear system} (ELS). This is effective in 2D, where the proxy count scales linearly with the unit-cell boundary length and the ELS is small. It does not transfer cheaply to 3D: the proxy count scales quadratically, the ill-conditioned rectangular ELS block grows accordingly, and, because proxy strengths and wall densities are co-solved, the ELS is tied to the embedded surfaces and must be re-assembled when the geometry changes. Eliminating this cost is one of the design goals of the present scheme. Pei et al.~\cite{pei2023fast} address the same ELS bottleneck through a low-rank plane-wave factorization of the far field, accelerated by the non-uniform fast Fourier transform (NUFFT) \cite{barnett2019parallel} and insensitive to unit-cell aspect ratio; the construction requires a kernel-specific Sommerfeld-type derivation for each new kernel, and a three-dimensional realization has not been demonstrated.

The closest 3D precursor is the multipole-to-local (M2L) approach of Yan and Shelley~\cite{yan2018}, which combines a near/far splitting with kernel-independent FMM (KIFMM) for the unbounded $N$-body sum in singly-, doubly-, and triply-periodic boxes. The far-field M2L operator there is built from the far-field portion of the {\em periodic} Green's function, constructed via Ewald summation; the splitting parameter must be chosen aggressively to meet the target precomputation accuracy, inflating the near-field to several layers of image boxes. Being aimed at the unbounded $N$-body sum, \cite{yan2018} does not couple directly to a surface-BIE solver, so wall-bounded partial periodicity is not a primary use case.

The present scheme retains the free-space-Green's-function philosophy of \cite{marple2016fast, barnett2018unified} and the KIFMM near/far splitting of \cite{yan2018}, but combines them so as to avoid the ELS solve. The KIFMM equivalent surface around the central period box serves as the auxiliary proxy basis, and the contribution from far image boxes is captured by a hierarchical proxy sum --- in the spirit of the lattice-sum renormalization of Berman and Greengard~\cite{berman1994renormalization}, but with kernel-independent equivalent sources in place of analytic multipoles --- rendered absolutely convergent by a net-force-zero compatibility condition. As a result: (a) no periodic Green's function or Ewald splitting is constructed; (b) the periodization precomputation depends only on the periodic-box geometry, independent of the kernel and of the embedded surfaces, and is reused as-is across the Stokeslet, stresslet, and rotlet, replacing the per-geometry ELS solve of \cite{marple2016fast, barnett2018unified}; (c) the near-field uses a single layer of image boxes, rather than the multiple layers required by \cite{yan2018}; and (d) the periodization plugs directly into a free-space, high-order, adaptive surface-BIE infrastructure, so wall-bounded periodic geometries --- pipes with circular cross-section, slabs between parallel planes, and dense suspensions of arbitrarily shaped particles --- can be handled directly.

\Cref{sec:formulation} states the unified periodic BIE covering all three target geometries and the role of the background flow in absorbing the prescribed pressure drop. \Cref{sec:method} develops the periodization scheme in detail: the KIFMM proxy construction, the hierarchical summation over image boxes, and the precomputed operator that maps the upward proxy density to the local proxy density. \Cref{sec:numerical} validates the method through convergence studies against exact and manufactured solutions, parameter sensitivity tests for the multipole order and hierarchy depth, and strong and weak scaling experiments on polydisperse suspensions with up to $3.8\times10^{6}$ degrees of freedom. \Cref{sec:conclusions} summarizes the main contributions and outlines directions for future work.

\section{Mathematical formulation}
 \label{sec:formulation}
 \newcommand\Real{{\ensuremath{\mathbb{R}}}}
\newcommand\StokesSL{{\ensuremath{\mathcal{S}}}}
\newcommand\StokesDL{{\ensuremath{\mathcal{D}}}}

We consider an incompressible Newtonian fluid governed by the Stokes equations
\begin{align}
  -\Delta \boldsymbol{u} + \nabla p &= 0, \label{eq:stokes_momentum} \\
  \nabla \cdot \boldsymbol{u} &= 0, \label{eq:stokes_continuity}
\end{align}
in a (possibly periodic) fluid domain bounded by a smooth surface \(\Gamma\) on which a Dirichlet velocity \(\boldsymbol{u} = \boldsymbol{g}\) is prescribed. Throughout, \(\boldsymbol{n}\) denotes the unit normal on \(\Gamma\) pointing into the fluid. The three target geometries treated by the present method --- a triply-periodic unit cell, a wall-bounded doubly-periodic channel, and a singly-periodic channel (illustrated by representative examples in \cref{fig:all_periodicity_examples}, panels (c), (b), (a) respectively) --- are all formulated by a single boundary integral equation, parameterized by the periodic lattice and a background flow. We first recall the free-space layer potentials (\cref{sec:free_space_BIE}) and then state the unified periodic BIE (\cref{sec:periodic_unified}), with the three concrete cases listed in~\cref{tab:periodic_cases}.

\subsection{Free-space layer potentials and BIE}
\label{sec:free_space_BIE}

Let \(\boldsymbol{r} = \boldsymbol{x} - \boldsymbol{y}\) and \(r = \|\boldsymbol{r}\|\). The free-space Stokeslet \(\boldsymbol{S}\) (the fundamental solution of~\cref{eq:stokes_momentum,eq:stokes_continuity} for a point force) and the associated double-layer (traction) kernel \(\boldsymbol{T}(\,\cdot\,;\boldsymbol{n})\) are
\begin{equation}
  \boldsymbol{S}(\boldsymbol{x},\boldsymbol{y}) = \frac{1}{8\pi}\!\left(\frac{I}{r} + \frac{\boldsymbol{r}\boldsymbol{r}^{T}}{r^{3}}\right),
  \qquad
  \boldsymbol{T}(\boldsymbol{x},\boldsymbol{y};\boldsymbol{n}) = -\frac{3}{4\pi}\,\frac{\boldsymbol{r}\boldsymbol{r}^{T}\,(\boldsymbol{r}\cdot\boldsymbol{n})}{r^{5}},
\end{equation}
where \(\boldsymbol{T}\) is the rank-three Stokes stresslet contracted with the unit normal \(\boldsymbol{n}\). They generate the Stokes single- and double-layer potentials,
\begin{equation}
  \StokesSL[\bm{\mu}](\boldsymbol{x}) = \int_{\Gamma}\!\boldsymbol{S}(\boldsymbol{x},\boldsymbol{y})\,\bm{\mu}(\boldsymbol{y})\,d\Gamma(\boldsymbol{y}),
  \qquad
  \StokesDL[\bm{\mu}](\boldsymbol{x}) = \int_{\Gamma}\!\boldsymbol{T}(\boldsymbol{x},\boldsymbol{y};\boldsymbol{n}_{\boldsymbol{y}})\,\bm{\mu}(\boldsymbol{y})\,d\Gamma(\boldsymbol{y}),
  \label{eq:layer_potentials}
\end{equation}
both of which satisfy~\cref{eq:stokes_momentum,eq:stokes_continuity} away from \(\Gamma\) for any sufficiently smooth density \(\bm{\mu}\) on \(\Gamma\). For the standard free-space Dirichlet problems, taking the principal-value limit \(\boldsymbol{x}\to\Gamma\) and applying the jump relations yields the classical second-kind boundary integral equations
\begin{align}
  \text{Interior \(\Omega\subset\Real^3\):}\quad &
  \left(\tfrac12 I + \StokesDL\right)\![\bm{\mu}](\boldsymbol{x}) = \boldsymbol{g}(\boldsymbol{x}), \\
  \text{Exterior, decay at \(\infty\):}\quad &
  \left(\tfrac12 I + \StokesDL + \StokesSL\right)\![\bm{\mu}](\boldsymbol{x}) = \boldsymbol{g}(\boldsymbol{x}),
\end{align}
on \(\boldsymbol{x}\in\Gamma\); the single-layer term in the exterior representation supplies the net-force degree of freedom that the double layer alone cannot represent~\cite{pozrikidis1992boundary, youngren1975stokes}. The interior problem requires the compatibility condition \(\int_{\Gamma}\boldsymbol{g}\cdot\boldsymbol{n}\,d\Gamma = 0\); the exterior problem (with \(\boldsymbol{u}\to\boldsymbol{0}\) at infinity) is uniquely solvable for any smooth \(\boldsymbol{g}\) on \(\Gamma\).

\subsection{A unified periodic BIE formulation}
\label{sec:periodic_unified}

Let \(\mathcal{L}\subset\Real^3\) be a \(d\)-dimensional Bravais lattice with \(d\in\{1,2,3\}\), and let \(\Pi\subset\Real^3\) denote a primitive cell. We seek a velocity \(\boldsymbol{u}\) and pressure \(p\) satisfying~\cref{eq:stokes_momentum,eq:stokes_continuity} in the periodic fluid domain
\[
  E_{\mathcal{L}} \;=\; \bigl(\Pi \setminus \overline{\Omega}\bigr) + \mathcal{L},
\]
where \(\Omega\) is the union of any confining walls and embedded particles inside \(\Pi\) and \(\Gamma=\partial\Omega\) carries the prescribed velocity \(\boldsymbol{g}\). The flow is required to be lattice-periodic in velocity and to admit a prescribed pressure drop in each periodic direction:
\begin{align}
  \boldsymbol{u}(\boldsymbol{x}+\boldsymbol{\ell}) &= \boldsymbol{u}(\boldsymbol{x}),
  &
  p(\boldsymbol{x}+\boldsymbol{\ell}) &= p(\boldsymbol{x}) - p^{\text{drop}}_{\boldsymbol{\ell}},
  &&\boldsymbol{\ell}\in\mathcal{L},
  \label{eq:periodicity}
\end{align}
where \(p^{\text{drop}}_{\boldsymbol{\ell}}\) is the pressure drop across one period in the direction of \(\boldsymbol{\ell}\) (so \(p^{\text{drop}}_{\boldsymbol{\ell}} > 0\) drives flow in the direction of \(\boldsymbol{\ell}\)). The boundary data must satisfy the compatibility condition \(\int_{\Gamma}\boldsymbol{g}\cdot\boldsymbol{n}\,d\Gamma = 0\). For wall-bounded geometries (channel between parallel planes; pipe with circular cross-section), \(\boldsymbol{u}\) vanishes on the confining walls, which are part of \(\Gamma\); for the wall-free triply-periodic case, \(\Gamma\) is the union of the embedded particle surfaces.

\paragraph{Background flow}
Let \(\boldsymbol{u}_{bg}\) be a known closed-form Stokes flow on \(E_{\mathcal{L}}\) that carries the prescribed pressure drop and is lattice-periodic in velocity.
\mydel{, and respects the no-slip condition on any confining wall.} %
Concrete choices for the three target geometries are listed in~\cref{tab:periodic_cases}. The disturbance \(\boldsymbol{u}-\boldsymbol{u}_{bg}\) is then lattice-periodic in both velocity and pressure (no pressure drop) and is represented by surface densities supported on \(\Gamma\).

\paragraph{Integral representation}
We represent the disturbance velocity by a periodized combined layer potential plus a constant offset \(\overline{\bm{\mu}}\):
\begin{equation}
  \boldsymbol{u}(\boldsymbol{x})
  \;=\; \boldsymbol{u}_{bg}(\boldsymbol{x})
  \;+\; \overline{\bm{\mu}}
  \;+\; \sum_{\boldsymbol{\ell}\in\mathcal{L}}
        (\StokesDL + \StokesSL)\bigl[\bm{\mu}-\overline{\bm{\mu}}\bigr](\boldsymbol{x}-\boldsymbol{\ell}),
  \qquad \boldsymbol{x}\in E_{\mathcal{L}},
  \label{eq:unified_rep}
\end{equation}
where \(\bm{\mu}\) is the unknown surface density on \(\Gamma\) and
\(\overline{\bm{\mu}} \,=\, |\Gamma|^{-1}\!\int_{\Gamma}\bm{\mu}\,d\Gamma\)
is its mean. Subtracting \(\overline{\bm{\mu}}\) inside the lattice sum imposes the zero-mean condition
\(\int_{\Gamma}(\bm{\mu}-\overline{\bm{\mu}})\,d\Gamma = \boldsymbol{0}\)
implicitly; this is the compatibility condition required for the lattice sum to converge absolutely~\cite{hasimoto1959periodic}. Adding \(\overline{\bm{\mu}}\) outside the sum supplies the rigid translational degree of freedom the constraint would otherwise remove.

Taking the principal-value limit \(\boldsymbol{x}\to\Gamma\) from the fluid side and applying the jump relations gives the master boundary integral equation
\begin{equation}
  \overline{\bm{\mu}}
  \;+\; \sum_{\boldsymbol{\ell}\in\mathcal{L}}
        \!\left(\tfrac12 I + \StokesDL + \StokesSL\right)\!\bigl[\bm{\mu}-\overline{\bm{\mu}}\bigr](\boldsymbol{x}-\boldsymbol{\ell})
  \;=\; \boldsymbol{g}(\boldsymbol{x}) \,-\, \boldsymbol{u}_{bg}(\boldsymbol{x}),
  \qquad \boldsymbol{x}\in\Gamma,
  \label{eq:unified_bie}
\end{equation}
a second-kind Fredholm equation for \(\bm{\mu}\).
\mydel{A uniform background flow \(\boldsymbol{U}_{\!\infty}\) at zero pressure drop is incorporated by setting \(\boldsymbol{u}_{bg} = \boldsymbol{U}_{\!\infty}\) and \(\boldsymbol{p}^{\text{drop}}=\boldsymbol{0}\); the right-hand side is then \(\boldsymbol{g}-\boldsymbol{U}_{\!\infty}\).} %

\paragraph{Specialization to the three target geometries}
\Cref{eq:unified_rep,eq:unified_bie} apply verbatim to all three cases. The cases differ only in the lattice \(\mathcal{L}\), the primitive cell \(\Pi\), and the closed-form background flow \(\boldsymbol{u}_{bg}\), summarized in~\cref{tab:periodic_cases}.

\begin{table}[ht]
\centering
{\fboxsep=12pt%
\colorbox{black!4}{%
\begin{minipage}{0.95\textwidth}
\centering
\small
\renewcommand{\arraystretch}{1.5}
\begin{tabular*}{\linewidth}{@{\extracolsep{\fill}} p{0.2\textwidth} c c p{0.55\textwidth} @{}}
\toprule
Geometry & \(d\) & Lattice \(\mathcal{L}\) & Background flow \(\boldsymbol{u}_{bg}(\boldsymbol{x})\) \\
\midrule
\textbf{Triply-periodic suspension} \newline{\scriptsize\rule{0pt}{10pt} \(\Pi=[0,L]^{3}\); \(\Gamma\) = particle surfaces; \cref{fig:all_periodicity_examples}(c)} &
3 &
\(L\,\mathbb{Z}^{3}\) &
\(\displaystyle\sum_{\boldsymbol{\ell}\in\mathcal{L}}\,\!\Bigl(\StokesSL\!\Bigl[\dfrac{\boldsymbol{p}^{\text{drop}} L^{2}}{|\Gamma|}\Bigr]\!(\boldsymbol{x}-\boldsymbol{\ell}) - \int_{\Pi}\!\boldsymbol{S}(\boldsymbol{x}-\boldsymbol{\ell},\boldsymbol{y})\,\dfrac{\boldsymbol{p}^{\text{drop}}}{L}\,d\boldsymbol{y}\Bigr)\)
\newline{\scriptsize\rule{0pt}{10pt} uniform-body-force flow producing the prescribed drop; the volume integral enforces zero net force per cell; equivalently $\int_{\Pi+\boldsymbol{\ell}}\boldsymbol{S}(\boldsymbol{x},\boldsymbol{y})\,d\boldsymbol{y}$ after $\boldsymbol{y}\mapsto\boldsymbol{y}+\boldsymbol{\ell}$} \\
\addlinespace
\textbf{Wall-bounded doubly-periodic channel} \newline{\scriptsize\rule{0pt}{10pt} \(\Pi=[0,L]^{2}\!\times[0,H]\); walls at \(z=0,H\); \cref{fig:all_periodicity_examples}(b)} &
2 &
\(L\,\mathbb{Z}^{2}\!\times\!\{0\}\) &
\(\displaystyle \frac{z(H-z)}{2L}\,\bigl(p_{1}^{\text{drop}},\,p_{2}^{\text{drop}},\,0\bigr)\)
\newline{\scriptsize\rule{0pt}{10pt} planar Poiseuille} \\
\addlinespace
\textbf{Singly-periodic pipe with circular cross-section} \newline{\scriptsize\rule{0pt}{10pt} axis along \(\boldsymbol{e}_1\), period \(L\); cross-sectional radius \(r(s)\) may vary smoothly along the pipe provided it is periodic, with inlet/outlet value \(R\)} &
1 &
\(L\,\mathbb{Z}\!\times\!\{0\}^{2}\) &
\(\displaystyle -\frac{p_{1}^{\text{drop}}}{4L}\,\bigl(R^{2}-y^{2}-z^{2},\,0,\,0\bigr)\)
\newline{\scriptsize\rule{0pt}{10pt} Hagen--Poiseuille flow for a straight circular cylinder of radius \(R\); satisfies no-slip on the pipe wall exactly only when \(r\equiv R\), otherwise the BIE disturbance corrects for the discrepancy} \\
\bottomrule
\end{tabular*}
\end{minipage}%
}}%
\caption{Specialization of the unified periodic formulation~\cref{eq:unified_rep,eq:unified_bie} to the three target geometries. In every case \(\boldsymbol{p}^{\text{drop}}=(p_{1}^{\text{drop}},p_{2}^{\text{drop}},p_{3}^{\text{drop}})\) is the prescribed pressure drop per period, with components in non-periodic directions set to zero.}
\label{tab:periodic_cases}
\end{table}

\paragraph{Force balance}
The constraint \(\int_{\Gamma}(\bm{\mu}-\overline{\bm{\mu}})\,d\Gamma = \boldsymbol{0}\) does \emph{not} mean that the fluid exerts zero net force on \(\Gamma\). In wall-bounded cases (rows 2 and 3 in~\cref{tab:periodic_cases}), the net pressure-drop force is delivered entirely by the background flow \(\boldsymbol{u}_{bg}\), which is an exact Stokes solution carrying the prescribed gradient. The layer-potential disturbance therefore carries no net force --- precisely the zero-mean condition. The drag on each interior particle follows from integrating the disturbance traction over that particle's surface. In the triply-periodic case (row 1), the volume integral in \(\boldsymbol{u}_{bg}\) plays the same role: it inserts a uniform body force whose total cancels that of the constant single-layer term, so that the periodic representation is force-free over \(\Pi\) while the imposed pressure gradient is realized exactly~\cite{hasimoto1959periodic}.

\paragraph{Wall-bounded geometries as a special case of triply-periodic}
Rows 2 and 3 of~\cref{tab:periodic_cases} may equivalently be treated as instances of the triply-periodic formulation: one extends the no-slip condition into the region outside the channel/pipe and treats the resulting box as triply-periodic, since the solution outside the channel/pipe is irrelevant. The advantage of the partial-periodicity formulation given here is that the corresponding \(\boldsymbol{u}_{bg}\) has a closed form in the prescribed pressure drop, so the right-hand side of~\cref{eq:unified_bie} is known analytically and the unknown density is supported only on the physical surfaces \(\Gamma\).

\section{Periodization scheme}
 \label{sec:method}
 \newcommand{\drawSquareWithBoundaryNodes}[8]{
  \draw[line width=#7, color=#6] (#1,#2) rectangle (#3,#4);

  \foreach \i in {0,...,#5} { %
    \fill[#6] ({#1 + (\i/(#5))*abs(#3-#1)}, #2) circle (#8);
  };
  \foreach \i in {1,...,#5} { %
    \fill[#6] (#3, {#2 + (\i/(#5))*abs(#4-#2)}) circle (#8);
  };
  \foreach \i in {1,...,#5} { %
    \fill[#6] ({#3 - (\i/(#5))*abs(#3-#1)}, #4) circle (#8);
  };
  \foreach \i in {1,...,#5} { %
    \fill[#6] (#1, {#4 - (\i/(#5))*abs(#4-#2)}) circle (#8);
  };
}

\newcommand{\Xup}{\ensuremath{\boldsymbol{x}^{\text{u}}}} %
\newcommand{\Xdn}{\ensuremath{\boldsymbol{x}^{\text{d}}}} %
\newcommand{\Yup}{\ensuremath{\boldsymbol{y}^{\text{u}}}} %
\newcommand{\Ydn}{\ensuremath{\boldsymbol{y}^{\text{d}}}} %
\newcommand{\UpDen}{\ensuremath{\boldsymbol{\mu}^{\text{u}}}} %
\newcommand{\DnDen}{\ensuremath{\boldsymbol{\mu}^{\text{d}}}} %
\newcommand{\Uup}{\ensuremath{\boldsymbol{u}^{\text{u}}}} %
\newcommand{\Udn}{\ensuremath{\boldsymbol{u}^{\text{d}}}} %
\newcommand{\Nproxy}{\ensuremath{N_{\text{proxy}}}} %
\newcommand{\Ncheck}{\ensuremath{N_{\text{check}}}} %

\newcommand{\Mmm}{\ensuremath{\mathsf{M}_{\text{M2M}}}}
\newcommand{\Mll}{\ensuremath{\mathsf{M}_{\text{L2L}}}}
\newcommand{\Mml}{\ensuremath{\mathsf{M}_{\text{M2L}}}}
\newcommand{\Pequiv}{\ensuremath{\mathsf{P}_{\bm{\mu}}}}
\newcommand{\Pcheck}{\ensuremath{\mathsf{P}_{\bm{u}}}}

We describe the periodization scheme that converts the free-space FMM into an efficient solver for the periodic boundary integral equation \cref{eq:unified_bie}.
The key idea is to split the infinite lattice sum into a near-field part, handled directly by the FMM with periodicity-aware interaction lists, and a far-field part, compressed into a compact set of proxy sources precomputed once and reused across solves.

Consider the layer potential operator on a boundary $\Gamma$ in free space:
\begin{equation}
  \boldsymbol{u}(\boldsymbol{x})
  = \int_\Gamma G(\boldsymbol{x}, \boldsymbol{y}) \, \bm{\mu}(\boldsymbol{y}) \, d\Gamma(\boldsymbol{y}),
  \label{e:freeSpace}
\end{equation}
where $\boldsymbol{x}$ is the target point,
$G(\boldsymbol{x}, \boldsymbol{y})$ is the free-space Green's function of the PDE (standing generically for any of the Stokes kernels: Stokeslet $\boldsymbol{S}$, stresslet $\boldsymbol{T}$, or the rotlet),
and $\bm{\mu}(\boldsymbol{y})$ is a density function supported on $\Gamma$.
We discretize~\cref{e:freeSpace} using the locally corrected Nystr\"om method~\cite{Malhotra2024}:
\begin{equation}
  \boldsymbol{u}(\boldsymbol{x}) \approx
  \sum_{k=1}^{N_s} G(\boldsymbol{x}, \boldsymbol{y}_k) \, \bm{\mu}(\boldsymbol{y}_k) \, w_k
  + \sum_{\boldsymbol{y}_k \in \mathcal{N}(\boldsymbol{x})} \boldsymbol{E}_k(\boldsymbol{x}) \, \bm{\mu}(\boldsymbol{y}_k),
  \label{e:discreteFreeSpace}
\end{equation}
where $\{\boldsymbol{y}_k\}$ and $\{w_k\}$ are the quadrature nodes and weights, respectively.
The correction coefficients $\boldsymbol{E}_k(\boldsymbol{x})$ are nonzero only in a small neighborhood $\mathcal{N}(\boldsymbol{x})$ of the target, ensuring high-order accuracy even when $\boldsymbol{x}$ is close to $\Gamma$.
The first term in~\cref{e:discreteFreeSpace} is an $N$-body interaction, which can be accelerated using the FMM.

We periodize the layer potential operator by considering contributions from infinitely many periodic copies of the boundary:
\begin{equation}
  \boldsymbol{u}(\boldsymbol{x})
  = \sum_{\boldsymbol{p} \in \mathbb{Z}^d}
  \int_\Gamma G\big(\boldsymbol{x}, \boldsymbol{y}+g(\boldsymbol{p})\big)
  \, \bm{\mu}(\boldsymbol{y}) \, d\Gamma(\boldsymbol{y}),
  \label{e:periodicSum}
\end{equation}
where $d$ is the dimension of periodicity and $g(\boldsymbol{p})$ is the translation vector associated with the periodic image indexed by $\boldsymbol{p} \in \mathbb{Z}^d$, so that $g(\boldsymbol{p}) \in \mathcal{L}$ in the lattice notation of \cref{sec:formulation} (e.g., $g(\boldsymbol{p})=L\boldsymbol{p}$ for a triply-periodic box of side $L$); cf.~\eqref{periodicSum} in \cref{sec:intro}.
The resulting potential is periodic up to an affine drift: it splits into a periodic part \(\boldsymbol{u}_p\) and an affine field
\(\boldsymbol{u}_0(\boldsymbol{x}) = \boldsymbol{c}_0 + \sum_{i=1}^d \boldsymbol{c}_i\,x_i\)
with vector coefficients \(\boldsymbol{c}_0,\boldsymbol{c}_i\in\mathbb{R}^3\):
\begin{equation}
  \boldsymbol{u}(\boldsymbol{x}) = \boldsymbol{u}_p(\boldsymbol{x}) + \boldsymbol{u}_0(\boldsymbol{x}),
  \qquad
  \boldsymbol{u}_p(\boldsymbol{x}+g(\boldsymbol{p})) = \boldsymbol{u}_p(\boldsymbol{x}),
  \qquad \boldsymbol{p} \in \mathbb{Z}^d.
  \label{e:periodic_with_drift}
\end{equation}
We similarly discretize \cref{e:periodicSum} using the locally corrected Nystr\"om scheme:
\begin{equation}
  \boldsymbol{u}(\boldsymbol{x}) \approx
  \sum_{\boldsymbol{p} \in \mathbb{Z}^d} \sum_{k=1}^{N_s}
  G\big(\boldsymbol{x}, \boldsymbol{y}_k+g(\boldsymbol{p})\big)
  \, \bm{\mu}(\boldsymbol{y}_k) \, w_k
  + \sum_{\boldsymbol{y}_k \in \mathcal{N}_{g}(\boldsymbol{x})}
  \boldsymbol{E}_k(\boldsymbol{x}) \, \bm{\mu}(\boldsymbol{y}_k),
  \label{e:discretePeriodicSum}
\end{equation}
where the first term involves an infinite sum over all periodic images of the source box shifted by \(g(\boldsymbol{p})\).
The second term is a local correction analogous to that in~\cref{e:discreteFreeSpace}, but with $\mathcal{N}_g(\boldsymbol{x})$ enlarged to include quadrature nodes from nearby periodic images.
This guarantees accuracy even when the target \(\boldsymbol{x}\) is close to the boundary of the periodic cell.
In the remainder of this section, we describe the periodic FMM used to evaluate the first term in \cref{e:discretePeriodicSum}.

\subsection{Periodic fast multipole method}
The FMM computes the $N$-body sum for a set of $N$ sources and $N$ targets in a cubic box in $\mathcal{O}(N)$ time.
It works by partitioning the domain using a hierarchical tree structure and computing interactions between well-separated groups of sources and targets organized hierarchically into tree nodes.
These interactions are approximated using multipole and local expansions that approximate the far field of a group of sources and the incoming field at a group of targets, respectively.
In this work, we use a version of the FMM based on the proxy points called the Kernel Independent FMM~\cite{Ying2004KIFMM,Malhotra2015,Malhotra2016}.
Below, we describe how to extend this method to compute the potential from an infinite periodic array of sources.

\begin{figure}[htbp]
  \centering
  \resizebox{0.9\linewidth}{!}{\begin{tikzpicture}[x=1.0cm,y=1.0cm]

    \begin{scope}[shift={(-20,1.5)}, scale=5]
      \fill[pattern color=red!70,pattern=north east lines] (-0.4,-0.4) rectangle (3.4,3.4);
      \draw[step=1,red!70,line width=5pt] (-0.4,-0.4) grid (3.4,3.4);
      \fill[white] (0,0) rectangle ++(3,3);

      \fill[white] (1,1) rectangle ++(1,1);
      \draw[step=1,black!20,line width=5pt] (0,0) grid (3,3);

      \draw[red!70,line width=5pt] (0,0) rectangle (3,3);
      \draw[black!99,line width=5pt] (1,1) rectangle (2,2);
      \node at (1.5,1.5) {\fontsize{36}{38}\selectfont \bf{B}};

      \node at (1.5,2.7) {\Huge \bf Proxy sources};
      \node at (1.5,2.45) {\Huge \bf $(\Ydn,\DnDen)$};

      \drawSquareWithBoundaryNodes{0.15}{0.15}{2.85}{2.85}{6}{orange}{5pt}{1.2pt}; %
    \end{scope}
    \draw[dash pattern=on 10pt off 5pt,line width=3pt] (-15,6.5) -- (0,0);
    \draw[dash pattern=on 10pt off 5pt,line width=3pt] (-15,11.5) -- (0,16);
    \draw[dash pattern=on 10pt off 5pt,line width=3pt] (-10,6.5) -- (16,0);
    \draw[dash pattern=on 10pt off 5pt,line width=3pt] (-10,11.5) -- (16,16);
    \fill[white] (0,0) rectangle ++(16,16);

    \fill[blue!40] (6,8) rectangle ++(6,6);
    \fill[blue!60] (7,9) rectangle ++(3,3);
    \node at (8.6,10.5) {\Huge $\bf n_1$};

    \fill[red!40] (14,2) rectangle ++(2,6);
    \fill[red!40] (0,2) rectangle ++(4,6);
    \fill[red!60] (15,3) rectangle ++(1,3);
    \fill[red!60] (0,3) rectangle ++(2,3);
    \node at (0.6,4.5) {\Huge $\bf{n_2}$};

    \draw[step=1,black!60,line width=0.4pt] (0,0) grid (16,16);
    \draw[step=2,black!80,line width=1.4pt] (0,0) grid (16,16);
    \draw[step=4,black!99,line width=3.0pt] (0,0) grid (16,16);
    \draw[step=8,black!99,line width=6.0pt] (0,0) grid (16,16);
    \draw[black!99,line width=6.0pt] (0,0) rectangle (16,16);

  \end{tikzpicture}}
  \caption{
    Left: A cubic domain B tiled periodically in 2D.
    The contributions from image boxes beyond the nearest neighbors are approximated using proxy sources (orange) placed around the target box.
    Right: The domain B is partitioned hierarchically using a tree structure (a quadtree in 2D) for the FMM.
    For two tree nodes \(n_1\) and \(n_2\), the shaded regions show their FMM interaction lists.
    For a node \(n_1\), in the interior of the domain, the interaction list consists of boxes in its neighborhood.
    For a node \(n_2\), near the left edge of the domain, the interaction list also includes boxes on the opposite side of the domain due to periodicity.
    This captures the contributions from sources in the first layer of neighboring image boxes around the original source box B.
  }
  \label{fig:fmm-periodization}
\end{figure}
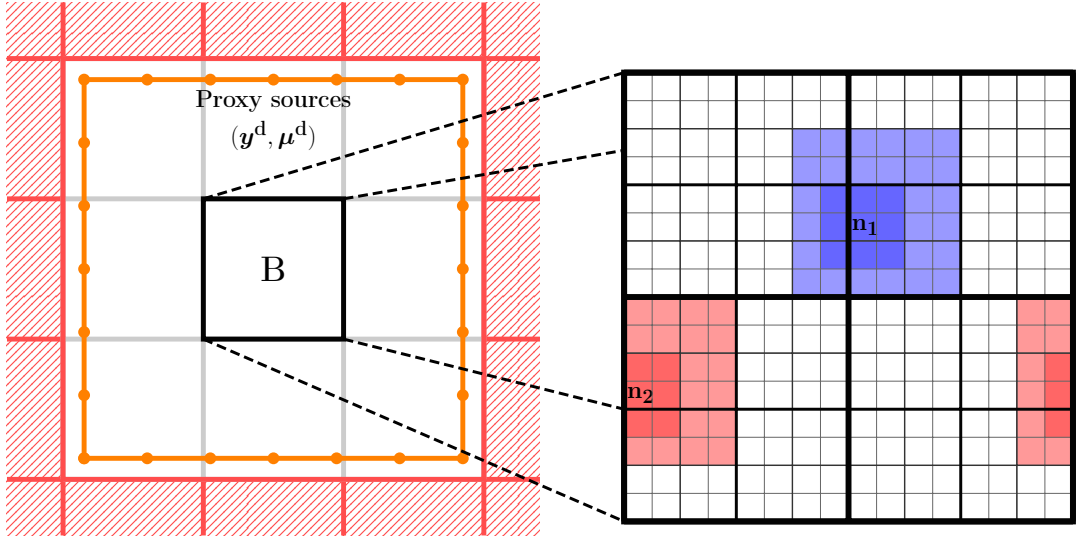

Consider the following infinite sum of periodic images of a set of \(N_s\) sources in a cubic box \(B\):
\begin{equation}
  \boldsymbol{u}(\boldsymbol{x}) =
  \sum_{\boldsymbol{p} \in\mathbb{Z}^d} \sum_{k=1}^{N_s} G(\boldsymbol{x},\boldsymbol{y}_k+g(\boldsymbol{p})) \, \bm{\mu}_k,
  \label{e:periodicNbody}
\end{equation}
where \(\boldsymbol{y}_k\) are the source locations inside box \(B\),
and \(\bm{\mu}_k\) are their strengths.
We evaluate this infinite sum by explicitly including just one layer of neighboring images around the original source box
and approximate all remaining distant images with a compact set of proxy sources placed around the domain (\cref{fig:fmm-periodization}-left):
\begin{equation}
  \boldsymbol{u}(\boldsymbol{x}) \approx
  \sum_{\substack{\boldsymbol{p} \in \{-1,0,1\}^d \\ \boldsymbol{p} \neq \boldsymbol{0}}}
  \sum_{k=1}^{N_s} G(\boldsymbol{x},\boldsymbol{y}_k+g(\boldsymbol{p})) \, \bm{\mu}_k
  + \sum_{j=1}^{N_{\text{proxy}}} G(\boldsymbol{x},\Ydn_j) \, \DnDen_j,
  \label{e:periodicNbodyProxy}
\end{equation}
where \(\Ydn_j\) are the proxy source locations and \(\DnDen_j\) their corresponding strengths.
The first term in \cref{e:periodicNbodyProxy} can be computed using a standard FMM by creating copies of the original box.
However, this adds redundant work of constructing the multipole expansions for each identical copy.
Instead, we modify the FMM to account for periodicity directly in the tree structure
by modifying the interaction lists of tree nodes near the box boundary to include contributions from tree nodes on the opposite side of the box (\cref{fig:fmm-periodization}-right).
The second term in \cref{e:periodicNbodyProxy} captures the far field of all remaining periodic images.
Below, we describe how to compute the proxy source strengths \(\DnDen_j\).

\begin{table}[t]
  \centering
  \begin{tabular}{ll}
    \toprule
    Symbol & Description \\
    \midrule
    \(\Yup, \UpDen\) & proxy point locations and strengths for outgoing (multipole) expansions \\
    \(\Xup, \Uup\) & check point locations and potentials for outgoing (multipole) expansions \\
    \(\Ydn, \DnDen\) & proxy point locations and strengths for incoming (local) expansions \\
    \(\Xdn, \Udn\) & check point locations and potentials for incoming (local) expansions \\
    \(\Nproxy\) & number of proxy points \\
    \(\Ncheck\) & number of check points \\
    \bottomrule
  \end{tabular}
  \caption{Notation for proxy and check points used in the Kernel Independent FMM.}
  \label{t:kifmm-notation}
\end{table}

\subsubsection{Multipole expansions via proxy points \label{ss:multipole}}
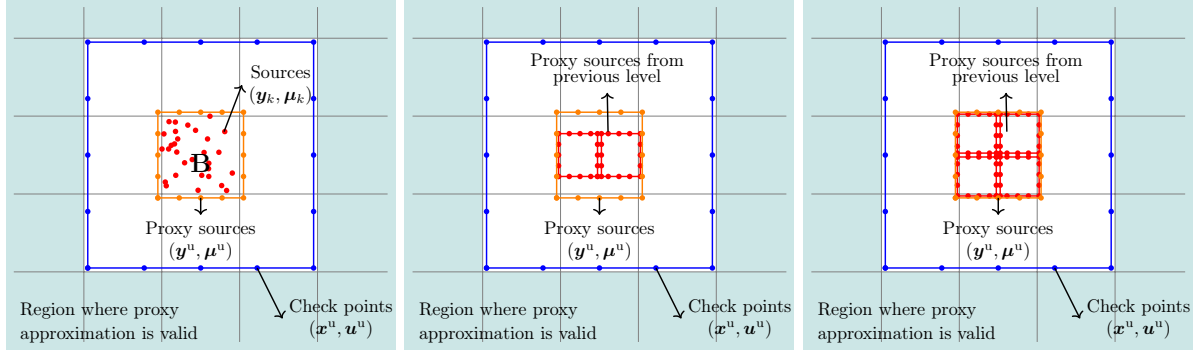
\begin{figure}[htbp]
  \resizebox{0.33\linewidth}{!}{\begin{tikzpicture}[scale=1.5]

    \fill[teal!20] (-3,-3) rectangle (2,1.5);
    \fill[white] (-2,-2) rectangle (1,1);

    \draw[step=1cm,black!50,very thin] (-2.9,-2.9) grid (1.9,1.4);

    \foreach \i in {1,...,30} {
      \fill[red] ({-1.0 + (rand*0.5+0.5)*1.0}, {-1.0 + (rand*0.5+0.5)*1.0}) circle (1pt);
    };

    \drawSquareWithBoundaryNodes{-1.05}{-1.05}{0.05}{0.05}{4}{orange}{0.8pt}{1.0pt};

    \drawSquareWithBoundaryNodes{-1.95}{-1.95}{0.95}{0.95}{4}{blue}{0.8pt}{1.0pt};

    \draw[->, thick] (-0.2,-0.2) -- (0.03,0.4) node[right, align=center] {Sources \\ $(\boldsymbol{y}_k,\bm{\mu}_k)$};
    \draw[->, thick] (-0.5,-1.05) -- (-0.5,-1.27) node[below, align=center] {Proxy sources \\ $(\Yup,\UpDen)$};
    \draw[->, thick] (0.225,-1.95) -- (0.55,-2.6) node[right, align=center] {Check points \\ $(\Xup,\Uup)$};
    \node[right] at (-2.9, -2.5) {Region where proxy};
    \node[right] at (-2.9, -2.8) {approximation is valid};
    \node at (-0.5,-0.6) {\Large \bf{B}};

  \end{tikzpicture}}
  \resizebox{0.33\linewidth}{!}{\begin{tikzpicture}[scale=1.5]

    \fill[teal!20] (-3,-3) rectangle (2,1.5);
    \fill[white] (-2,-2) rectangle (1,1);

    \draw[step=1cm,black!50,very thin] (-2.9,-2.9) grid (1.9,1.4);

    \drawSquareWithBoundaryNodes{-0.525}{-0.775}{0.025}{-0.225}{4}{red}{0.8pt}{1.0pt};
    \drawSquareWithBoundaryNodes{-1.025}{-0.775}{-0.475}{-0.225}{4}{red}{0.8pt}{1.0pt};

    \drawSquareWithBoundaryNodes{-1.05}{-1.05}{0.05}{0.05}{4}{orange}{0.8pt}{1.0pt};

    \drawSquareWithBoundaryNodes{-1.95}{-1.95}{0.95}{0.95}{4}{blue}{0.8pt}{1.0pt};

    \node[above] at (-0.4,0.52) {Proxy sources from};
    \draw[->, thick] (-0.39,-0.2) -- (-0.4,0.3) node[above] {previous level};

    \draw[->, thick] (-0.5,-1.05) -- (-0.5,-1.27) node[below, align=center] {Proxy sources \\ $(\Yup,\UpDen)$};
    \draw[->, thick] (0.225,-1.95) -- (0.55,-2.6) node[right, align=center] {Check points \\ $(\Xup,\Uup)$};
    \node[right] at (-2.9, -2.5) {Region where proxy};
    \node[right] at (-2.9, -2.8) {approximation is valid};

  \end{tikzpicture}}
  \resizebox{0.33\linewidth}{!}{\begin{tikzpicture}[scale=1.5]

    \fill[teal!20] (-3,-3) rectangle (2,1.5);
    \fill[white] (-2,-2) rectangle (1,1);

    \draw[step=1cm,black!50,very thin] (-2.9,-2.9) grid (1.9,1.4);

    \drawSquareWithBoundaryNodes{-0.525}{-0.525}{0.025}{0.025}{4}{red}{0.8pt}{1.0pt};
    \drawSquareWithBoundaryNodes{-1.025}{-0.525}{-0.475}{0.025}{4}{red}{0.8pt}{1.0pt};
    \drawSquareWithBoundaryNodes{-0.525}{-1.025}{0.025}{-0.475}{4}{red}{0.8pt}{1.0pt};
    \drawSquareWithBoundaryNodes{-1.025}{-1.025}{-0.475}{-0.475}{4}{red}{0.8pt}{1.0pt};

    \drawSquareWithBoundaryNodes{-1.05}{-1.05}{0.05}{0.05}{4}{orange}{0.8pt}{1.0pt};

    \drawSquareWithBoundaryNodes{-1.95}{-1.95}{0.95}{0.95}{4}{blue}{0.8pt}{1.0pt};

    \node[above] at (-0.4,0.52) {Proxy sources from};
    \draw[->, thick] (-0.39,-0.2) -- (-0.4,0.3) node[above] {previous level};

    \draw[->, thick] (-0.5,-1.05) -- (-0.5,-1.27) node[below, align=center] {Proxy sources \\ $(\Yup,\UpDen)$};
    \draw[->, thick] (0.225,-1.95) -- (0.55,-2.6) node[right, align=center] {Check points \\ $(\Xup,\Uup)$};
    \node[right] at (-2.9, -2.5) {Region where proxy};
    \node[right] at (-2.9, -2.8) {approximation is valid};

  \end{tikzpicture}}
  \caption{
    Left: Construction of a multipole expansion for a source box.
    The potential is first evaluated at check points (blue) due to the physical sources (red) inside the box.
    Proxy sources (orange) are then placed on the box boundary with strengths chosen to reproduce the check-point potentials.
    The resulting representation is valid outside the check surface, in the so-called admissible region.
    Center: The same procedure can be applied hierarchically.
    Here, two expansions from the previous level are combined into a single expansion that captures their joint far field.
    Repeating this process produces multipole expansions for increasingly large collections of source boxes tiled in 1D, 2D, or 3D.
    If level~0 corresponds to the original box, then level~$l$ represents $2^l$ copies tiled in each dimension.
    Right: Computation of the multipole expansion for a 2D tiling of four source boxes.
  }
  \label{fig:s2m-m2m}
\end{figure}

We revisit the construction of multipole expansions using the proxy point method~\cite{Ying2004KIFMM}.
We summarize the notation for proxy and check points in \cref{t:kifmm-notation}.
As illustrated in \cref{fig:s2m-m2m}-left, we can approximate the far field of
a set of sources in a box B using a set of proxy sources placed on the box boundary.
First, we evaluate the potential at a set of check points (blue) due to the physical sources (red) inside the box:
\begin{equation}
  \Uup_i = \sum_{k=1}^{N_s} G(\Xup_i, \boldsymbol{y}_k) \, \bm{\mu}(\boldsymbol{y}_k),
  \quad i = 1, \ldots, \Ncheck,
\end{equation}
where $\Xup_i$ are the check point locations, $\boldsymbol{y}_k$ are the source locations, and
$\bm{\mu}(\boldsymbol{y}_k)$ are their strengths.
Next, we solve for the strengths of proxy sources (orange) on the box boundary so that they reproduce the potential at the check points:
\begin{equation}
  \Uup_i = \sum_{j=1}^{\Nproxy} G(\Xup_i, \Yup_j) \, \UpDen_j,
  \quad i = 1, \ldots, \Ncheck,
  \label{e:multipole_solve}
\end{equation}
where $\Yup_j$ are the proxy source positions and $\UpDen_j$ their strengths.
The resulting proxy representation is valid outside the check surface (the admissible region), where it faithfully
approximates the far field of the original sources.
The accuracy of this approximation is controlled by the number of check and proxy points.

In 3D, the box B is a cube, and the check and proxy points are placed on the boundaries of cubes surrounding B.
The points are on a uniform grid on each face of the cube, and the number of points per edge is called the multipole order \(m\).
The linear system in \cref{e:multipole_solve} is highly ill-conditioned;
therefore, the solution must be computed in a backward-stable manner to avoid large numerical errors~\cite{Malhotra2015}.

\subsubsection{Multipole-to-multipole translations}
The procedure in \cref{ss:multipole} can be used to construct the combined multipole expansion for a collection of source boxes whose multipole expansions have been previously computed.
As shown in~\cref{fig:s2m-m2m}-center, two adjacent boxes can be merged into a single expansion capturing their combined far field.
By applying this process hierarchically, we can build multipole expansions for increasingly large collections of the original box tiled in one, two, or three dimensions.
If level~0 corresponds to the original box, then level~$l$ corresponds to $2^l$ copies tiled in each periodic dimension.
We encode this merge step as a single \emph{multipole-to-multipole (M2M)} operator matrix $\Mmm$ that maps the proxy density at level~$l$ to the proxy density at level~$l+1$:
\begin{equation}
  \UpDen_{l+1} \;=\; \Mmm \, \UpDen_l,
  \qquad l = 0, 1, 2, \ldots.
  \label{e:M2M}
\end{equation}
For scale-invariant kernels, such as the Stokes kernels, this is the same operator \(\Mmm\) at every level.

The FMM constructs the multipole expansion of the root node B during its upward pass, giving the proxy point representation for its far field.
From this proxy point representation of the root node, we can construct the multipole expansions for a hierarchy of tilings of the original root box in 1D, 2D, or 3D.

\subsubsection{Multipole-to-local translations \label{ss:m2l}}
For each level $l$, we define the \emph{multipole-to-local (M2L)} operator
$\mathsf{M}_{\text{M2L},\,l}$ that maps the multipole proxy density $\UpDen_l$ to the
incoming check potential $\Udn_l$ on the level-$l$ check surface. It sums kernel
contributions from the level-$l$ image boxes that are well-separated from the
target but lie within its parent's neighborhood (\cref{fig:2d-m2l-l2l}-left):
\begin{equation}
  \Udn_l \;=\; \mathsf{M}_{\text{M2L},\,l} \, \UpDen_l,
  \qquad
  \bigl(\mathsf{M}_{\text{M2L},\,l}\bigr)_{ij}
  \;\coloneqq\;
  \sum_{\substack{\boldsymbol{p} \in \{-2,\dots,3\}^d \\ \|\boldsymbol{p}\|_\infty > 1}}
  G\!\left(\Xdn_{l,i},\, \Yup_{l,j} + 2^l\boldsymbol{p}\right).
  \label{e:M2L_level_l}
\end{equation}
The index set $\boldsymbol{p} \in \{-2,\dots,3\}^d$ with $\|\boldsymbol{p}\|_\infty > 1$ enumerates
the $6^d - 3^d$ children of the parent's $3^d$ colleagues, excluding the $3^d$ self-children that
are handled at the next-finer level.

By the scale invariance of $G$, the M2L operator at level $l$ is obtained from the level-$0$
operator $\Mml \coloneqq \mathsf{M}_{\text{M2L},\,0}$ by a pair of diagonal scaling matrices
$\mathsf{S}_{\text{c}}$ and $\mathsf{S}_{\text{p}}$ acting on the check and proxy sides,
respectively:
\begin{equation}
  \mathsf{M}_{\text{M2L},\,l} \;=\; \mathsf{S}_{\text{c}}^l \, \Mml \, \mathsf{S}_{\text{p}}^l.
  \label{e:M2L_scaling}
\end{equation}

\begin{figure}[htbp]
  \centering
  \resizebox{0.45\linewidth}{!}{\begin{tikzpicture}[scale=1.0]
    \useasboundingbox (-2.75,-2.75) rectangle (4.75,4.75);
    \clip (-2.75,-2.75) rectangle (4.75,4.75);

    \fill[pattern color=red!30,pattern=north east lines] (-2,-2) rectangle (4,4);
    \fill[white] (-1,-1) rectangle (2,2);
    \draw[step=1cm,black!30,very thin] (-4,-4) grid (8,8);

    \draw[line width=0.95pt, color=black] (0,0) rectangle (1,1);
    \node at (0.5,0.5) {\huge \bf B};
    \node[below,align=center] at (0.5,0) {\small $(\Xdn_0,\Udn_0)$};

    \foreach \x in {-2,...,3} { %
      \foreach \y in {-2,...,3} {
        \ifthenelse{\x<-1 \OR \x >1 \OR \y<-1 \OR \y>1}{
          \drawSquareWithBoundaryNodes{\x}{\y}{\x+1}{\y+1}{4}{red}{0.8pt}{1.0pt};
        }{}
      }
    };

    \node at (-0.3,-1.5) {\large \bf level 0};

    \drawSquareWithBoundaryNodes{-0.08}{-0.08}{1.08}{1.08}{4}{blue}{1.0pt}{1.5pt};
    \drawSquareWithBoundaryNodes{-0.92}{-0.92}{1.92}{1.92}{4}{orange}{1.0pt}{1.5pt};
    \node[align=center] at (1.68,0.5) {\small $\Ydn_0$ \\ $\DnDen_0$};

    \draw[-Latex, line width=1.0pt] (-1.50,-1.50) -- (-0.1,-0.1);
    \draw[-Latex, line width=1.0pt] ( 2.50,-1.50) -- ( 1.1,-0.1);
    \draw[-Latex, line width=1.0pt] (-1.50, 2.50) -- (-0.1, 1.1);
    \draw[-Latex, line width=1.0pt] ( 2.50, 2.50) -- ( 1.1, 1.1);
    \node at (0.08,1.6) {\small $\mathsf{M}_{\text{M2L},\,0}$};

  \end{tikzpicture}}
  \hfill
  \resizebox{0.45\linewidth}{!}{\begin{tikzpicture}[scale=1.0]
    \useasboundingbox (-2.75,-2.75) rectangle (4.75,4.75);
    \clip (-2.75,-2.75) rectangle (4.75,4.75);

    \fill[pattern color=red!30,pattern=north east lines] (-4,-4) rectangle (8,8);
    \fill[white] (-1,-1) rectangle (2,2);
    \fill[white!50] (-2,-2) rectangle (4,4);
    \draw[step=1cm,black!30,very thin] (-4,-4) grid (8,8);

    \draw[line width=0.95pt, color=black] (0,0) rectangle (1,1);
    \node at (0.5,0.5) {\huge \bf B};
    \node[above,align=center] at (0.5,1.0) {\small $(\Xdn_0,\Udn_0)$};
    \node[above,align=center] at (1.0,2.1) {\small $(\Xdn_1,\Udn_1)$};
    \node[above,align=center] at (1.0,-1.75) {\small $(\Ydn_1,\DnDen_1)$};

    \foreach \x in {-2,...,3} { %
      \foreach \y in {-2,...,3} {
        \ifthenelse{\x<-1 \OR \x >1 \OR \y<-1 \OR \y>1}{
          \drawSquareWithBoundaryNodes{\x*2}{\y*2}{\x*2+2}{\y*2+2}{4}{red}{0.8pt}{1.5pt};
        }{}
      }
    };

    \node at (-1.0,-2.4) {\Large \bf level 1};

    \drawSquareWithBoundaryNodes{-0.16}{-0.16}{2.16}{2.16}{4}{blue}{1.0pt}{1.5pt};
    \drawSquareWithBoundaryNodes{-0.08}{-0.08}{1.08}{1.08}{4}{blue}{1.0pt}{1.5pt};
    \drawSquareWithBoundaryNodes{-1.7}{-1.7}{3.7}{3.7}{4}{orange}{1.0pt}{1.5pt};

    \draw[-Latex, line width=1.2pt] (-2.50,-1.50) -- (-0.2,-0.2);
    \draw[-Latex, line width=1.2pt] ( 4.50,-1.50) -- ( 2.2,-0.2);
    \draw[-Latex, line width=1.2pt] (-2.50, 3.50) -- (-0.2, 2.2);
    \draw[-Latex, line width=1.2pt] ( 4.50, 3.50) -- ( 2.2, 2.2);
    \node at (-0.6,2.9) {\small $\mathsf{M}_{\text{M2L},\,1}$};

    \node[above] at (1.7,0.4) {\small $\Mll$};
    \draw[-Latex, line width=0.8pt] ( 2.16, 0.4) -- ( 1.08, 0.4);

  \end{tikzpicture}}
  \caption{
    Left: M2L translation at level $0$. The operator $\mathsf{M}_{\text{M2L},\,0}$ evaluates the potential $\Udn_0$ at
    the incoming check points $\Xdn_0$ around the root box B from the level-$0$ image copies of the multipole proxy
    sources (shown in red), as given by~\cref{e:M2L_level_l}.
    Right: M2L translation at level $1$, followed by L2L translation to level $0$. The operator
    $\mathsf{M}_{\text{M2L},\,1}$ evaluates the potential $\Udn_1$ at the level-$1$ incoming check points $\Xdn_1$; the
    L2L operator $\Mll$ then evaluates this potential to the level-$0$ incoming check points $\Xdn_0$. }
  \label{fig:2d-m2l-l2l}
\end{figure}
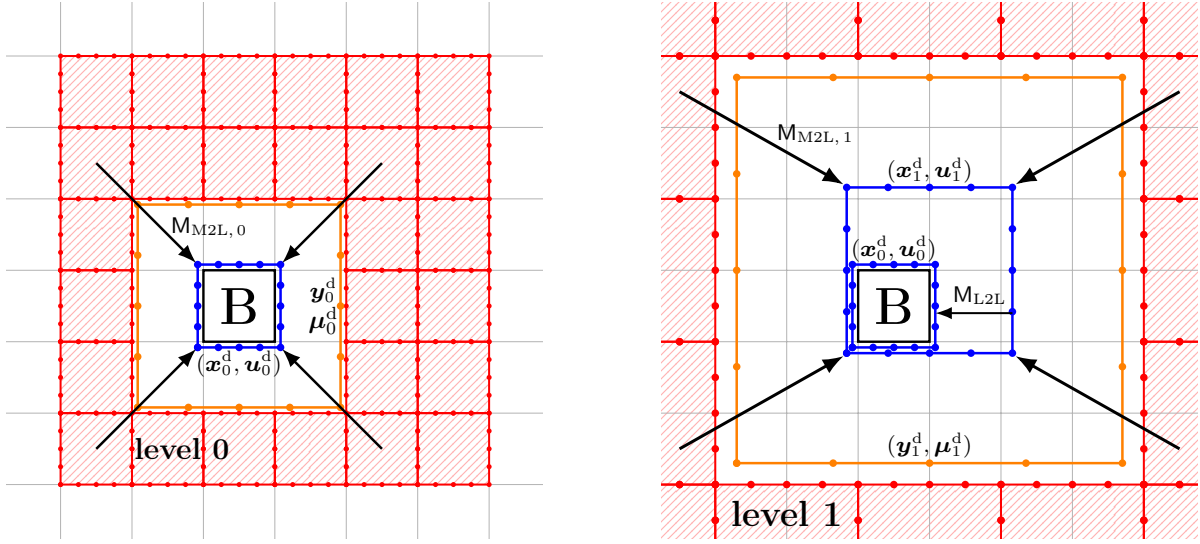

\begin{figure}[htbp]
  \centering
  \resizebox{0.98\linewidth}{!}{\begin{tikzpicture}[scale=0.9]
    \useasboundingbox (-6.5,-2.3) rectangle (10.5,4.0);
    \clip (-6.5,-2.3) rectangle (10.5,4.0);

    \draw[line width=0.95pt, color=black] (0,0) rectangle (1,1);
    \node at (0.5,0.5) {\large \bf B};

    \fill[pattern color=red,pattern=north east lines] (-8,0) rectangle (-1,1);
    \fill[pattern color=red,pattern=north east lines] (2,0) rectangle (16,1);

    \draw[step=1cm,black] (-8,0) grid (-1,1);
    \draw[step=1cm,black] (2,0) grid (16,1);

    \drawSquareWithBoundaryNodes{-2}{0}{-1}{1}{4}{red}{1.0pt}{1.5pt};
    \drawSquareWithBoundaryNodes{2}{0}{3}{1}{4}{red}{1.0pt}{1.5pt};
    \drawSquareWithBoundaryNodes{3}{0}{4}{1}{4}{red}{1.0pt}{1.5pt};
    \node[below] at (-1.45,0.0) {\scriptsize \bf level\,0};
    \node[below] at ( 3.00,0.0) {\scriptsize \bf level\,0};

    \drawSquareWithBoundaryNodes{-4}{-0.5}{-2}{1.5}{4}{red}{1.0pt}{1.5pt};
    \drawSquareWithBoundaryNodes{4}{-0.5}{6}{1.5}{4}{red}{1.0pt}{1.5pt};
    \drawSquareWithBoundaryNodes{6}{-0.5}{8}{1.5}{4}{red}{1.0pt}{1.5pt};
    \node[below] at (-3,-0.5) {\bf level 1};
    \node[below] at ( 6,-0.5) {\bf level 1};

    \drawSquareWithBoundaryNodes{-8}{-1.5}{-4}{2.5}{4}{red}{1.0pt}{1.5pt};
    \drawSquareWithBoundaryNodes{8}{-1.5}{12}{2.5}{4}{red}{1.0pt}{1.5pt};
    \drawSquareWithBoundaryNodes{12}{-1.5}{16}{2.5}{4}{red}{1.0pt}{1.5pt};
    \node[above] at (-5,-1.5) {\large \bf level 2};
    \node[above] at ( 9,-1.5) {\large \bf level 2};

    \drawSquareWithBoundaryNodes{-0.08}{-0.08}{1.08}{1.08}{4}{blue}{1.0pt}{1.5pt};
    \node[below,align=center] at (0.5,0.02) {\tiny $(\Xdn_0,\Udn_0)$};

    \drawSquareWithBoundaryNodes{-0.16}{-0.66}{2.16}{1.66}{4}{blue}{1.0pt}{1.5pt};
    \node[below,align=center] at (0.5,-0.57) {\small $(\Xdn_1,\Udn_1)$};

    \drawSquareWithBoundaryNodes{-0.24}{-1.74}{4.24}{2.74}{4}{blue}{1.0pt}{1.5pt};
    \node[below,align=center] at (0.5,-1.65) {\normalsize $(\Xdn_2,\Udn_2)$};

    \draw[-Latex, thick] (-1.5,1.05) to[out=40,in=140] node[pos=0.15, above, font=\footnotesize, yshift=1pt] {$\mathsf{M}_{\text{M2L},\,0}$} (0.2,1.11);
    \draw[-Latex, thick] (-3.0,1.55) to[out=40,in=140] node[pos=0.25, above, font=\normalsize, xshift=-2pt] {$\mathsf{M}_{\text{M2L},\,1}$} (0.7,1.69);
    \draw[-Latex, thick] (-5.0,2.55) to[out=40,in=140] node[pos=0.2, above, font=\large, xshift=-6pt] {$\mathsf{M}_{\text{M2L},\,2}$} (1.4,2.75);

    \draw[-Latex, thick] (2.8,1.05) to[out=140,in=40] node[pos=0.05, above, font=\footnotesize, xshift=5pt] {$\mathsf{M}_{\text{M2L},\,0}$} (0.8,1.11);
    \draw[-Latex, thick] (5.4,1.55) to[out=140,in=40] node[pos=0.1, above, font=\normalsize, xshift=6pt] {$\mathsf{M}_{\text{M2L},\,1}$} (1.3,1.69);
    \draw[-Latex, thick] (9.0,2.55) to[out=140,in=40] node[pos=0.2, above, font=\large, xshift=3pt] {$\mathsf{M}_{\text{M2L},\,2}$} (2.6,2.75);

    \draw[-Latex, thick] (1.37,-0.64) to node[pos=0.4, font=\footnotesize, xshift=1.15em] {$\Mll$} (1.08,-0.06);

    \draw[-Latex, thick] (2.21,-1.75) to node[font=\normalsize, xshift=1.3em] {$\Mll$} (1.77,-0.63);

    \end{tikzpicture}}
  \caption{
    Hierarchical computation of the periodic far field for 1D periodization.
    Once the multipole expansions have been constructed for each level, we compute the M2L interaction $\mathsf{M}_{\text{M2L},\,l}$ at each level $l$, evaluating the contribution of the level-$l$ image boxes (red) on the level-$l$ check surface (blue).
    We then apply the L2L interactions $\Mll$ to translate these check potentials down through the levels, finally arriving at the level-$0$ check points surrounding the target box B.
  }
  \label{fig:1d-periodization}
\end{figure}
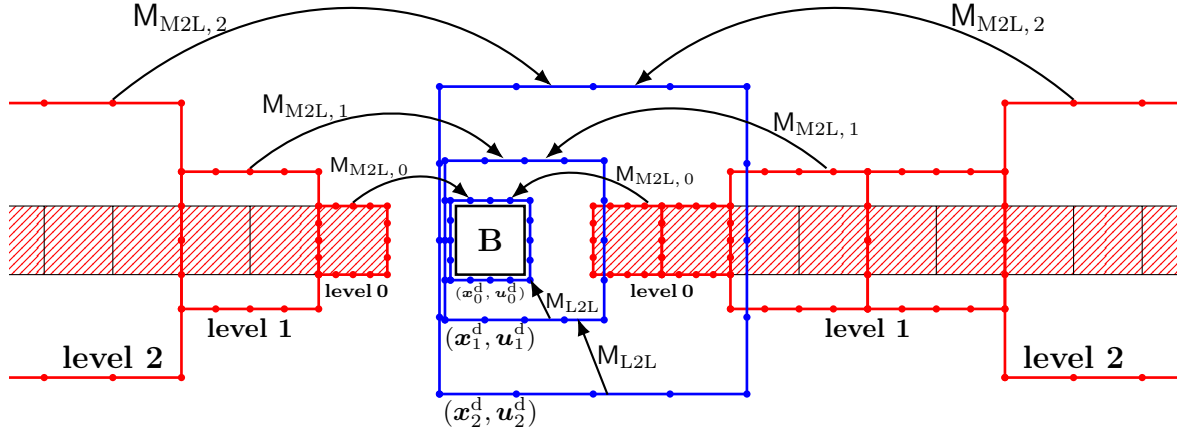

\subsubsection{Local-to-local translations}
At each level $l$, the M2L operator evaluates the potential $\Udn_l$ at the level-$l$ check points due to the level-$l$
image boxes.
The \emph{local-to-local (L2L)} operator $\Mll$ translates this potential to the level-$(l-1)$ check points by first
solving for the level-$l$ incoming proxy density $\DnDen_l$ that reproduces $\Udn_l$ on the level-$l$ check surface, and
then evaluating the resulting expansion on the level-$(l-1)$ check surface (\cref{fig:2d-m2l-l2l}-right):
\begin{equation}
  \Udn_{l-1} \;=\; \Mll \, \Udn_l.
  \label{e:L2L}
\end{equation}
As with the multipole construction, the proxy solve from $\Udn_l$ to $\DnDen_l$ is ill-conditioned and must be computed
in a backward-stable manner.
For scale-invariant kernels, this is the same operator \(\Mll\) at every level.

\subsubsection{Hierarchical summation of periodic images \label{ss:hierarchical_sum}}
With the translation operators defined above, we can build the operator matrix that computes the hierarchical sum of the
potential from well-separated images of the original root box at the incoming check points of the root box:
\begin{equation}
  \mathsf{M}_{N_\text{levels}} \coloneqq
  \sum_{l=0}^{N_{\text{levels}}} \Mll^l \, \mathsf{S}_{\text{c}}^l \, \Mml \, \mathsf{S}_{\text{p}}^l \, \Mmm^l,
  \label{e:hierarchical_sum}
\end{equation}
or equivalently as the recursion
\begin{equation}
  \mathsf{M}_{n+1} \;=\; \Mml \;+\; \Mll \, \mathsf{S}_{\text{c}} \, \mathsf{M}_n \, \mathsf{S}_{\text{p}} \, \Mmm,
  \qquad n = 0, 1, \ldots, N_{\text{levels}}-1,
  \label{e:bare_recursion}
\end{equation}
with the base case $\mathsf{M}_0 = \Mml$.
The number of levels is chosen sufficiently large so that this sum has converged to the desired accuracy; we use $N_{\text{levels}}=45$ throughout.

\subsubsection{Stabilization of the recursion via projection operators \label{ss:stabilization}}
The infinite lattice sum is only convergent when the net force in the source box vanishes; the bare
recursion~\cref{e:bare_recursion} is therefore conditionally stable, and any nonzero-net-force component of $\UpDen_0$
introduced by round-off is amplified exponentially across the $N_{\text{levels}}$ iterations.
We project out this mode at every step with two operators:
$\Pequiv = I - \boldsymbol{e} \boldsymbol{1}^\top$, where $\boldsymbol{e}$ is the outgoing proxy density
corresponding to a uniform unit source over the root box;
and $\Pcheck = I - \boldsymbol{1} \boldsymbol{1}^\top/N_\text{check}$, which subtracts the per-kernel-component mean of
the check potential.
The stabilized recursion is
\begin{equation}
  \mathsf{M}_0 \;=\; \Pcheck \, \Mml \, \Pequiv,
  \qquad
  \mathsf{M}_{n+1} \;=\; \Pcheck \, \bigl(\Mml \;+\; \Mll \, \mathsf{S}_{\text{c}} \, \mathsf{M}_n \, \mathsf{S}_{\text{p}} \, \Mmm\bigr) \, \Pequiv,
  \label{e:stable_recursion}
\end{equation}
for $n = 0, \ldots, N_{\text{levels}}-1$.

\subsubsection{Corner correction for periodic continuity \label{ss:corner_correction}}
As noted earlier, the periodic sum results in solutions that are periodic up to an affine drift (\(\boldsymbol{u}_0\) in \cref{e:periodic_with_drift}).
We therefore have to correct for this affine drift in the periodic directions.
We fit the trilinear polynomial $\sum_{\alpha\in\{0,1\}^3} a_\alpha \boldsymbol{q}^\alpha$ to the check potential
evaluated at the eight corners of the unit cube%
\footnote{The potential at each corner is evaluated by applying the periodization operator $\mathsf{M}_{\text{BC}}$ and
adding M2L contributions from the image boxes not already included in $\mathsf{M}_{\text{BC}}$, excluding any image box
that touches that corner. The excluded region around each corner is the same; therefore, the same constant
value is excluded from each corner.}.
We retain the relevant monomial terms of this polynomial according to the periodicity and subtract it from the solution.
The relevant monomial terms are
$\{1, q^x, q^y, q^z\}$ for triple periodicity,
$\{1, q^x, q^y\}$ for double, and $\{1,
q^x\}$ for single periodicity.
Denoting the resulting correction matrix by $\mathsf{M}_{\text{corner}}$, the final lattice-sum operator is
\begin{equation}
  \mathsf{M}_{\text{BC}} \;=\; \mathsf{M}_{N_{\text{levels}}} \;-\; \mathsf{M}_{\text{corner}}.
  \label{e:bc_with_corner}
\end{equation}

\subsubsection{Proxy source representation for the incoming expansion \label{ss:proxy_solve_incoming}}
The root-box incoming check potential $\Udn_0$ due to all distant periodic images of the source box is given by
$\Udn_0 \;=\; \mathsf{M}_{\text{BC}} \UpDen_0$.
We then solve for the proxy source strengths $\DnDen_0$ around the target box that produce the same potential $\Udn_0$.
As before, this linear system must be solved in a backward-stable manner.
The resulting proxy source representation approximates the incoming potential due to all periodic copies of the original
source box, except for the nearest layer of images, which are handled separately by the FMM.
We precompute this operator matrix $\mathcal{P}$ that directly maps the proxy source density of the multipole expansion
of root node B to the proxy source density of the incoming expansion of B, such that:
\(\DnDen=\mathcal{P}\,\UpDen\).

\subsubsection{Summary of the FMM periodization scheme}
The periodization scheme operates in two phases.

\paragraph{Precomputation (once per box geometry)}
The operator matrices $\Mmm$, $\Mll$, $\Mml$, together with the per-level scaling matrices $\mathsf{S}_{\text{c}}, \mathsf{S}_{\text{p}}$, are built once from the level-$0$ proxy and check surfaces using the KIFMM machinery of \cref{ss:multipole}.
The projectors $\Pequiv$ acting on $\UpDen$ and $\Pcheck$ acting on $\Udn$ are constructed in \cref{ss:stabilization}.
The stabilized recursion~\cref{e:stable_recursion} is then iterated for $N_{\text{levels}}$ steps to produce the lattice-sum matrix $\mathsf{M}_{N_{\text{levels}}}$.
The corner correction $\mathsf{M}_{\text{corner}}$ of \cref{ss:corner_correction} is subtracted to obtain $\mathsf{M}_{\text{BC}}$, which is then composed with the backward-stable solve of~\cref{ss:proxy_solve_incoming} for the incoming proxy density to give the periodization operator $\mathcal{P}\!:\UpDen\mapsto\DnDen$, stored as a dense $\Nproxy \times \Nproxy$ matrix.
The operator $\mathcal{P}$ depends only on the periodic-box geometry (a unit cube here), the periodization (1D, 2D, or 3D) and the accuracy parameters (the multipole order $m$, recursion depth $N_{\text{levels}}$) and is reused verbatim across all subsequent solves and across the Stokeslet, stresslet, and rotlet kernels without modification.

\paragraph{Evaluation (each GMRES iteration)}
Each application of the periodized layer potential operator~\cref{eq:unified_bie} to the current density iterate proceeds as follows:
\begin{enumerate}
  \item The FMM upward pass computes \(\UpDen\) from the source distribution.
  \item The precomputed operator gives the far-field incoming proxy density: \(\DnDen = \mathcal{P}\,\UpDen\), at cost \(\mathcal{O}(\Nproxy^2)\).
  \item The FMM downward pass evaluates the far-field contribution at all targets, with the periodic interaction lists in the FMM tree handling near-field contributions from the first layer of image boxes.
  \item Locally corrected Nystr\"om quadrature corrections are added for target--source pairs in \(\mathcal{N}_g(\boldsymbol{x})\).
\end{enumerate}
The total per-iteration cost is \(\mathcal{O}(N)\), dominated by the FMM. The periodization matrix--vector product in step~2 costs \(\mathcal{O}(\Nproxy^2)=\mathcal{O}(m^4)\) per iteration (the one-time precomputation of $\mathcal{P}$ via the hierarchical sum costs $\mathcal{O}(N_{\text{levels}} \, m^6)$) and is negligible in practice.

\section{Numerical results}
\label{sec:numerical}
\usepgfplotslibrary{colorbrewer}
\usepgfplotslibrary{groupplots}
\pgfplotsset{compat=1.18}

In this section, we verify the convergence and performance of the scheme on a range of periodic geometries, from a simple cylindrical channel to a dense polydisperse suspension.
For all experiments, we apply a block-diagonal left-preconditioner: on a singly-periodic channel, we use the self-interaction matrix of a one-panel cylinder; on a particle, we use the self-interaction matrix of a sphere. 
The left-preconditioner needs to be computed only once per set of discretization parameters.

\paragraph{Notation}
Throughout this section, we report accuracy as the maximum relative error using a reference or exact solution $\ux_e(\x)$:
$$\text{Max.\ Relative Error}=\frac{\max_{\x}|\ux(\x)-\ux_e(\x)|}{\max_{\x}|\ux_e(\x)|}.$$
We report the setup time for building the boundary integral operator as $T_{\text{setup}}$,
and the time for one evaluation of the operator as $T_{\text{eval}}$.
The boundary integral operator is the combined field operator in \cref{eq:unified_bie},
requiring evaluation of the Stokes single- and double-layer operators.
In parallel runs $T_{\text{setup}}$ and $T_{\text{eval}}$ are wall-clock times.

\paragraph{Computing resources}
All experiments were performed on the Rusty cluster at the Flatiron Institute, which comprises 216 dual-socket compute nodes, each with two 32-core Intel Xeon Platinum 8362 CPUs and 1\,TB of memory, interconnected via InfiniBand.

\paragraph{Software}
Our software was run on a Linux operating system and compiled using \lstinline{GCC}-13.3.0, with the
\texttt{OpenMPI} library version 4.1.8, and optimization flags \lstinline[basicstyle=\ttfamily]{-O2 -march=x86_64}. We linked against the \texttt{Intel
MKL} library version 2024.2.2 and the \texttt{FFTW} library version 3.3.10.

\subsection{Surface discretization}
\label{sec:surface_disc}

All non-planar surfaces considered in this work---particles, cylindrical channel walls, and obstacle surfaces in the singly-periodic pipe geometry---are surfaces of revolution.  This is a deliberate design choice, not a geometric limitation: it allows us to exploit the convergent slender-body quadrature (CSBQ) framework of Malhotra and Barnett~\cite{Malhotra2024} for all singular and near-singular layer-potential evaluations.

Each such surface is described by a smooth centerline curve $\boldsymbol{x}_c(s)$, $s\in[0,1)$, together with a smoothly varying cross-sectional radius $r(s)$.  At each centerline point, two orthonormal vectors $\boldsymbol{e}_1(s)$ and $\boldsymbol{e}_2(s)$, both perpendicular to the tangent $d\boldsymbol{x}_c/ds$, span the cross-sectional plane.  Points on the surface are then parameterized by
\begin{equation}
  \boldsymbol{y}(s,\theta) \;=\; \boldsymbol{x}_c(s) + r(s)\bigl[\boldsymbol{e}_1(s)\cos\theta + \boldsymbol{e}_2(s)\sin\theta\bigr],
  \qquad s\in[0,1),\;\theta\in[0,2\pi).
  \label{eq:surface_param}
\end{equation}

The surface is discretized by partitioning the centerline arc into $N_p$ panels, each carrying $p = 10$ Chebyshev nodes in $s$, crossed with $N_f$ uniformly spaced nodes in $\theta$.  The surface density $\bm{\mu}$ is represented by its values at the resulting $N_p \times p \times N_f$ tensor-product nodes; for a vector-valued density in $\mathbb{R}^3$, the total number of degrees of freedom per body is $3\,N_p\, p\, N_f$.  Both parameters converge spectrally: exponentially in $\theta$ via the trapezoidal rule for smooth periodic functions, and at the Chebyshev rate in $s$ on each panel.

The CSBQ near-field and on-surface corrections exploit the rotational symmetry about the centerline: the angular $\theta$-integrals in the layer potentials reduce to modal (toroidal) Green's functions, for which precomputed generalized Chebyshev quadrature rules are used (see \cite{Malhotra2024}, Sections~3.2.2--3.2.4 for details). The CSBQ library leverages SCTL~\cite{SCTL} for its boundary integral framework and PVFMM~\cite{Malhotra2015} for its parallelized FMM. Visualizations were produced with ParaView.

Throughout the numerical experiments, $N_p$ denotes the number of centerline panels per body and $N_f$ the number of azimuthal Fourier modes.

\subsection{Self-convergence}
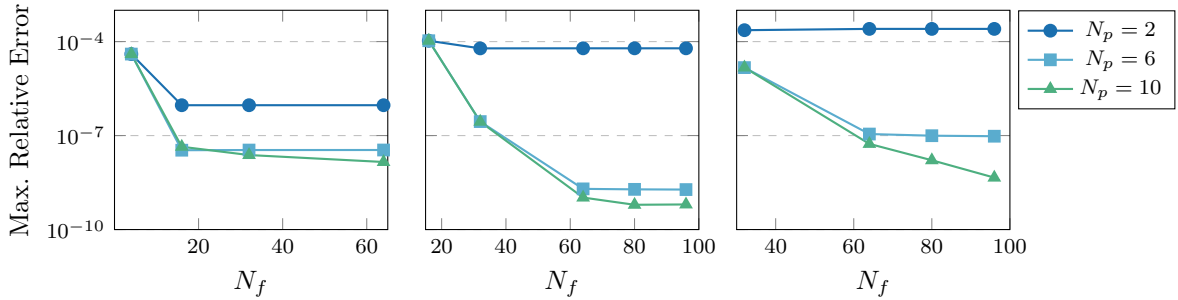
\begin{figure}[htbp]
\centering
\begin{tikzpicture}
\begin{groupplot}[
  group style={group size=3 by 1, horizontal sep=0.5cm},
  width=.33\linewidth,
  ymode=log,
  ymin=1e-10, ymax=1e-3,     %
  ymajorgrids=true,
  minor y tick num=8,
  grid style=dashed,
  ticklabel style={font=\footnotesize},
  legend style={font=\footnotesize}
]

\nextgroupplot[
  xlabel={$N_f$},
  ylabel={Max.\ Relative Error},
  xtick={20,40,60},
  xmin=0, xmax=65,
  legend pos=north east
]
\addplot[color={rgb,255:red,26;green,111;blue,175},mark=*,mark size=2.2pt,line width=0.8pt] coordinates {
    (4, 4.001198383e-05)(16, 9.314892949e-07)(32, 9.319589325e-07)(64, 9.321924752e-07)};
\addplot[color={rgb,255:red,90;green,172;blue,206},mark=square*,mark size=2.0pt,line width=0.8pt] coordinates {
    (4, 4.001338243e-05)(16, 3.413632896e-08)(32, 3.453908092e-08)(64, 3.464076561e-08)};
\addplot[color={rgb,255:red,77;green,175;blue,128},mark=triangle*,mark size=2.2pt,line width=0.8pt] coordinates {
    (4, 4.001582789e-05)(16, 4.39504899e-08)(32, 2.391510066e-08)(64, 1.434754892e-08)};

\nextgroupplot[
  xlabel={$N_f$},
  xtick={20,40,60,80,100},
  xmin=15, xmax=100,
  legend pos=outer north east,
  yticklabels={},           %
  ymajorgrids=true,
]
\addplot[color={rgb,255:red,26;green,111;blue,175},mark=*,mark size=2.2pt,line width=0.8pt] coordinates {
    (16, 1.063068679e-4)(32, 6.08e-5)(64, 6.11e-5)(80, 6.11e-5)(96, 6.11e-5)};
\addplot[color={rgb,255:red,90;green,172;blue,206},mark=square*,mark size=2.0pt,line width=0.8pt] coordinates {
    (16, 1.067497214e-4)(32, 2.80e-7)(64, 1.99e-9)(80, 1.92e-9)(96, 1.89e-9)};
\addplot[color={rgb,255:red,77;green,175;blue,128},mark=triangle*,mark size=2.2pt,line width=0.8pt] coordinates {
    (16, 1.067501075e-4)(32, 2.77e-7)(64, 1.05e-9)(80, 6.18e-10)(96, 6.33e-10)};

\nextgroupplot[
  xlabel={$N_f$},
  xtick={20,40,60,80,100},
  xmin=30, xmax=100,
  legend pos=outer north east,
  ymajorgrids=true,
  yticklabels={},
]
\addplot[color={rgb,255:red,26;green,111;blue,175},mark=*,mark size=2.2pt,line width=0.8pt] coordinates {
    (32, 0.000231452998)(64, 0.0002545642644)(80, 0.0002545301528)(96, 0.0002545015263)};
\addlegendentry{$N_p=2$}
\addplot[color={rgb,255:red,90;green,172;blue,206},mark=square*,mark size=2.0pt,line width=0.8pt] coordinates {
    (32, 1.48E-05)(64, 1.12E-07)(80, 9.91E-08)(96, 9.54E-08)};
\addlegendentry{$N_p=6$}
\addplot[color={rgb,255:red,77;green,175;blue,128},mark=triangle*,mark size=2.2pt,line width=0.8pt] coordinates {
    (32, 1.49E-05)(64, 5.43E-08)(80, 1.63E-08)(96, 4.55E-09)};
\addlegendentry{$N_p=10$}

\end{groupplot}
\end{tikzpicture}

\caption{Self-convergence in $N_f$ for singly-, doubly-, and triply-periodic pressure-driven flow past a spherical obstacle in a cylindrical channel (left), wall-confined toroidal loops (middle), and spherical suspensions (right), for several panel counts $N_p$. $N_p=18, N_f=96$ serves as the reference solution in each case. Spectral convergence is observed before plateaus caused by saturation of error from parameter $N_p$.}
\label{fig:self_conv_12peri}
\end{figure}

Using the singly-periodic interior Dirichlet formulation (\cref{sec:periodic_unified}), we solve for pressure-driven flow past cylindrical channels with or without periodic obstacles. The exact Poiseuille profile for an empty channel provides a validation benchmark. For a channel with a spherical obstacle, we investigate self-convergence in $N_f$ on both the channel and the obstacle in \cref{fig:self_conv_12peri}-left, at different $N_p$ on the obstacle, but fix the number of panels on the channel at $20$. The reference solution is chosen at obstacle panel $N_p=18$ and Fourier discretization $N_f=96$ on both channel and obstacle.

For the doubly-periodic formulation, we use the geometry of \cref{fig:all_periodicity_examples}(b) with $H=L=1$ with $p_1^{\text{drop}}=1$ and $p_2^{\text{drop}}=0$. Self-convergence for this configuration is shown in \cref{fig:self_conv_12peri}-middle. We fix the obstacles and target points not too close to the bounding planes so that standard Chebyshev quadrature is sufficiently accurate. We use flat bounding planes for simplicity, and to minimize the wall discretization error we simply split each of the top and bottom walls in the unit cell into a $2\times 2$ grid of panels, each with order-$49$ tensor-product Chebyshev nodes.

Finally, for triply-periodic flow, we vary the per-particle discretization on a 25-sphere polydisperse suspension under pressure-driven flow. We observe similar convergence at moderate parameters on each sphere in \cref{fig:self_conv_12peri}-right.

\subsection{Manufactured Solutions Test}

To validate the exterior Dirichlet formulation, we construct a manufactured solution
by placing a Stokes doublet (a pair of Stokeslets with equal and opposite forces)
inside each obstacle of a singly-periodic spherical suspension. Choosing doublets
(rather than single Stokeslets) ensures zero net force per period, satisfying the
compatibility condition required for absolute convergence of lattice sums.
We approximate the infinite lattice sum by
truncating to $2N_c+1$ cells along $\pm\bm{e}_1$ and take this truncation as the
reference solution $\ux_e(\x)$.

We first verify convergence of the truncated lattice sum. On a
uniform grid of target points in the unit cell, we measure the maximum relative error
between truncations at two values of $N_c$. For a polydisperse system of 25 spheres
(one doublet per sphere), $N_c = 6\times 10^4$ agrees with larger $N_c$ to within
$10^{-12}$, so we adopt it as the reference for all subsequent tests.

We use the reference solution evaluated on the boundary of the spheres as the boundary condition for our periodic solver to solve for the exterior field, then compare the solver solution with the reference solution to get the maximum relative error.
We use this error to get a heuristic accuracy estimate for the \textit{digits of accuracy} for a discretization parameter set $\{N_p, N_f\}$, which is the accuracy the parameter set should achieve based on its performance in this manufactured solution test.
\cref{tab:NpNf_digits} shows that only modest discretization is needed to achieve 3, 6, or 9 digits of accuracy. 

\begin{table}[H]
    \centering
    \renewcommand{\arraystretch}{1.2}
    \begin{tabular}{lccc}
        \toprule
        Digits of accuracy & 3 & 6 & 9 \\
        \midrule
        $N_p$ & 2 & 3 & 6 \\
        $N_f$ & 16 & 24 & 32 \\
        Max.\ Relative Error & $8.56\times10^{-5}$ & $2.20\times10^{-7}$ & $5.12\times10^{-10}$ \\
        \bottomrule
    \end{tabular}
    \caption{Number of panels ($N_p$) and Fourier modes ($N_f$) required to achieve at least 3, 6, or 9 digits of accuracy for the system of 25 spherical particles, measured against the manufactured solution.
    }
    \label{tab:NpNf_digits}
\end{table}

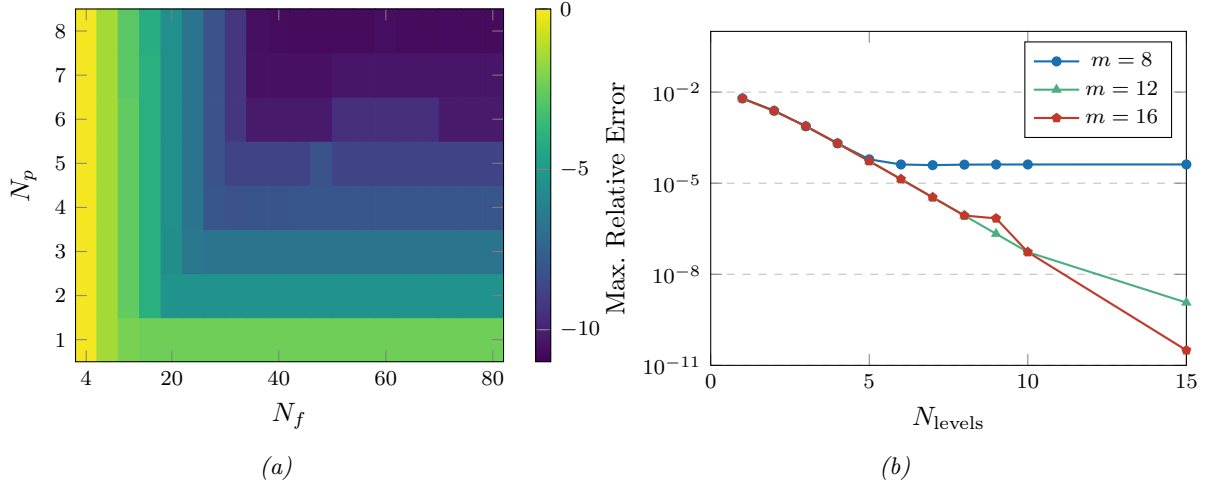
\begin{figure}[htbp]
\begin{subfigure}[b]{.46\textwidth}
    \begin{tikzpicture}
\begin{axis}[
    width=\linewidth,
    view={0}{90},
    xlabel={$N_f$},
    ylabel={$N_p$},
    colorbar,
    colormap/viridis,
    colorbar style={width=0.2cm},
    xmin=2, xmax=82,
    ymin=0.5, ymax=8.5,
    xtick={4, 20, 40, 60, 80},
    ytick={1, 2, 3, 4, 5, 6, 7, 8},
    ticklabel style={font=\footnotesize},
    point meta min=-11,
    point meta max=0,
    axis on top
]
    \addplot [
        matrix plot*,
        mesh/cols=20,
        mesh/rows=8
    ] table [point meta=\thisrow{z}]{
        x y z
        4 1 -0.137
        8 1 -1.502
        12 1 -2.225
        16 1 -2.399
        20 1 -2.408
        24 1 -2.409
        28 1 -2.409
        32 1 -2.409
        36 1 -2.409
        40 1 -2.409
        44 1 -2.409
        48 1 -2.409
        52 1 -2.409
        56 1 -2.409
        60 1 -2.409
        64 1 -2.409
        68 1 -2.408
        72 1 -2.408
        76 1 -2.408
        80 1 -2.408

        4 2 -0.138
        8 2 -1.492
        12 2 -2.668
        16 2 -4.067
        20 2 -5.386
        24 2 -5.384
        28 2 -5.385
        32 2 -5.384
        36 2 -5.384
        40 2 -5.384
        44 2 -5.384
        48 2 -5.384
        52 2 -5.384
        56 2 -5.384
        60 2 -5.384
        64 2 -5.384
        68 2 -5.384
        72 2 -5.384
        76 2 -5.384
        80 2 -5.384

        4 3 -0.138
        8 3 -1.492
        12 3 -2.668
        16 3 -4.068
        20 3 -5.577
        24 3 -6.658
        28 3 -6.764
        32 3 -6.764
        36 3 -6.767
        40 3 -6.764
        44 3 -6.764
        48 3 -6.764
        52 3 -6.764
        56 3 -6.764
        60 3 -6.764
        64 3 -6.764
        68 3 -6.764
        72 3 -6.764
        76 3 -6.764
        80 3 -6.764

        4 4 -0.138
        8 4 -1.492
        12 4 -2.667
        16 4 -4.068
        20 4 -5.565
        24 4 -6.777
        28 4 -8.093
        32 4 -8.149
        36 4 -8.143
        40 4 -8.143
        44 4 -8.089
        48 4 -8.089
        52 4 -8.089
        56 4 -8.089
        60 4 -8.089
        64 4 -8.089
        68 4 -8.089
        72 4 -8.089
        76 4 -8.089
        80 4 -8.089

        4 5 -0.138
        8 5 -1.492
        12 5 -2.668
        16 5 -4.067
        20 5 -5.565
        24 5 -6.777
        28 5 -8.169
        32 5 -8.793
        36 5 -8.777
        40 5 -8.775
        44 5 -8.775
        48 5 -8.169
        52 5 -8.775
        56 5 -8.775
        60 5 -8.775
        64 5 -8.775
        68 5 -8.775
        72 5 -8.775
        76 5 -8.775
        80 5 -8.775

        4 6 -0.138
        8 6 -1.492
        12 6 -2.668
        16 6 -4.067
        20 6 -5.565
        24 6 -6.777
        28 6 -8.168
        32 6 -9.291
        36 6 -10.250
        40 6 -10.236
        44 6 -10.237
        48 6 -10.236
        52 6 -9.510
        56 6 -9.510
        60 6 -9.510
        64 6 -9.510
        68 6 -9.510
        72 6 -10.232
        76 6 -10.233
        80 6 -10.234

        4 7 -0.138
        8 7 -1.492
        12 7 -2.822
        16 7 -4.067
        20 7 -5.565
        24 7 -6.777
        28 7 -8.168
        32 7 -9.289
        36 7 -10.553
        40 7 -10.551
        44 7 -10.554
        48 7 -10.553
        52 7 -10.379
        56 7 -10.377
        60 7 -10.379
        64 7 -10.378
        68 7 -10.379
        72 7 -10.380
        76 7 -10.380
        80 7 -10.380

        4 8 -0.138
        8 8 -1.492
        12 8 -2.822
        16 8 -4.191
        20 8 -5.565
        24 8 -6.777
        28 8 -8.168
        32 8 -9.289
        36 8 -10.602
        40 8 -10.662
        44 8 -10.712
        48 8 -10.712
        52 8 -10.710
        56 8 -10.710
        60 8 -10.616
        64 8 -10.708
        68 8 -10.712
        72 8 -10.708
        76 8 -10.710
        80 8 -10.710
    };
\end{axis}
\end{tikzpicture}
\caption{}
\label{fig:NpNf_param_grid}
\end{subfigure}\hfill
\begin{subfigure}[b]{.50\textwidth}
\begin{tikzpicture}
\begin{semilogyaxis}[
    xlabel={$N_{\text{levels}}$},
    ylabel={Max.\ Relative Error},
    xmin=0, xmax=15,
    ymin=1e-11, ymax=1,
    xtick={0,5,10,15},
    legend pos=north east,
    ymajorgrids=true,
    minor y tick num=8,
    grid style=dashed,
    ticklabel style={font=\footnotesize},
    legend style={font=\footnotesize},
    width=\linewidth,
    height=6cm
]
\addplot[color={rgb,255:red,26;green,111;blue,175},  mark=*,        mark size=1.5pt, line width=0.8pt]
    coordinates {
    (1,6.17E-03)(2,2.42E-03)(3,7.48E-04)(4,2.05E-04)(5,6.10E-05)
    (6,4.13E-05)(7,3.93E-05)(8,4.08E-05)(9,4.12E-05)(10,4.13E-05)
    (15,4.14E-05)(20,4.14E-05)};
\addlegendentry{$m=8$}
\addplot[color={rgb,255:red,77;green,175;blue,128},  mark=triangle*, mark size=1.5pt, line width=0.8pt]
    coordinates {
    (1,6.17E-03)(2,2.42E-03)(3,7.50E-04)(4,2.05E-04)(5,5.33E-05)
    (6,1.35E-05)(7,3.39E-06)(8,8.59E-07)(9,2.15E-07)(10,5.38E-08)
    (15,1.17E-09)(20,1.21E-09)};
\addlegendentry{$m=12$}
\addplot[color={rgb,255:red,192;green,57;blue,43},   mark=pentagon*, mark size=1.5pt, line width=0.8pt]
    coordinates {
    (1,6.17E-03)(2,2.42E-03)(3,7.50E-04)(4,2.05E-04)(5,5.34E-05)
    (6,1.36E-05)(7,3.42E-06)(8,8.59E-07)(9,6.93E-07)(10,5.38E-08)
    (15,3.14E-11)(20,6.31E-11)};
\addlegendentry{$m=16$}
\end{semilogyaxis}
\end{tikzpicture}
    \caption{}
    \label{fig:l_convergence}
\end{subfigure}
\caption{Convergence of the periodized solver for the 25-sphere singly-periodic system. (a)~Joint dependence of the maximum relative error on the number of Fourier modes $N_f$ and surface panels $N_p$, shown as $\log_{10}(\text{error})$. (b)~Convergence with respect to hierarchy depth $N_{\text{levels}}$ for three multipole orders $m$; higher $m$ is required to sustain convergence to smaller tolerances.}
\label{fig:convergence_combined}
\end{figure}

The solver converges exponentially in the number of azimuthal Fourier modes and to high order in the number of panels. Using the same manufactured solutions test, \cref{fig:NpNf_param_grid} illustrates this joint dependence for order-10 Chebyshev nodes.

The accuracy of the precomputed periodization operator depends additionally on the multipole order $m$ and the number of image levels $N_{\text{levels}}$. Fixing $N_p = 8$ and $N_f = 64$, \cref{fig:l_convergence} shows that the error decays exponentially with $N_{\text{levels}}$ for each multipole order. For $m=8$ the error saturates near $4\times10^{-5}$, where the multipole approximation error dominates before the image series has fully converged; the higher orders $m=12$ and $m=16$ continue to converge. We therefore fix $N_{\text{levels}}=45$ in practice, well beyond the onset of each plateau, and truncate \cref{fig:l_convergence} at $15$ levels to highlight the convergence regime. The multipole order is chosen to meet the prescribed layer-potential accuracy tolerance; for example, a tolerance of $10^{-14}$ requires $m=16$.

\subsection{Precomputation}
\begin{figure}[ht]
  \centering
    \begin{tikzpicture}
        \begin{loglogaxis}[
            width=0.5\linewidth,
            xlabel={$m$},
            ylabel={Time (s)},
            legend pos=north west,
            legend style={font=\footnotesize},
            ymajorgrids=true,
            grid style={dashed},
            log basis x={2},
            xtick={2,4,6,8,12,16},
            xticklabels={2,4,6,8,12,16}
        ]

        \addplot[color={rgb,255:red,26;green,111;blue,175}, mark=*, mark size=2.2pt, line width=0.8pt] coordinates { %
            ( 2,    1.0444)
            ( 4,    3.7585)
            ( 6,   15.2191)
            ( 8,   52.7508)
            (10,  158.8032)
            (12,  431.8158)
            (14, 1077.7527)
            (16, 2422.8584)
            (18, 4854.7701)
        };
        \addlegendentry{$\mathcal{S}$}

        \addplot[color=black, dashed, thick] coordinates {
            ( 5, 1.95577)
            (10, 125.1697)
            (12, 373.7549)
            (14, 942.4702)
            (16, 2100.0000)
            (18, 4257)
        };
        \addlegendentry{$\mathcal{O}(m^6)$}
        \end{loglogaxis}
    \end{tikzpicture}

\caption{Precomputation time for the Stokes single-layer ($\mathcal{S}$) periodizing operator as a function of multipole order $m$, using 64 OpenMP threads on a single Icelake compute node. $\mathcal{O}(m^6)$ is shown for reference at large $m$.}
\label{fig:precomp-time}
\end{figure}
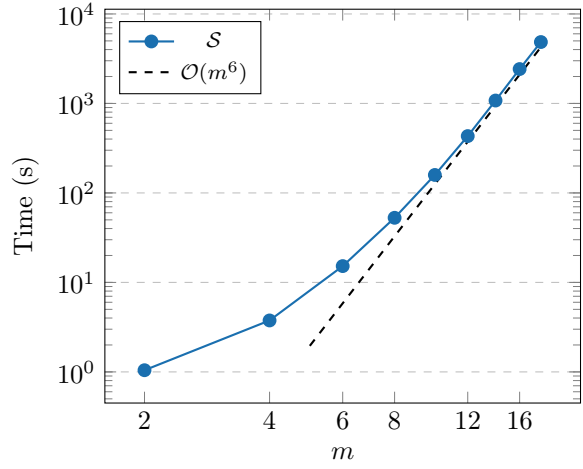

The periodization operator $\mathcal{P}$ is precomputed for different multipole orders $m$ depending on the user-defined accuracy tolerance, with $N_\text{levels}=45$ fixed, as described in \cref{sec:method}.
The precomputation cost scales as $\mathcal{O}(N_\text{proxy}^3) = \mathcal{O}(m^6)$ in the multipole order $m$.
In \cref{fig:precomp-time}, we show the precomputation time for the Stokes single-layer periodization operator as a function of $m$, demonstrating the expected $\mathcal{O}(m^6)$ scaling at large $m$.
The operator is stored as a dense $N_\text{proxy}\times N_\text{proxy}$ matrix.
Once stored, the operator can be applied verbatim across geometries and different kernels, resulting in a significant reduction in runtime.

\subsection{Periodization overhead\label{ss:periodization_overhead}}
In \cref{tab:periodization_time}, we compare the computational cost of a periodized solver against its free-space counterpart to quantify the overhead in the layer-potential operator setup and evaluation described in \cref{sec:method}.
The experimental setup is a polydisperse collection of spheres, each discretized with $N_p=6$ panels and $N_f=32$ Fourier modes, suspended in a Stokesian fluid.
We solve the triply-periodic problem with a prescribed pressure drop and the corresponding free-space problem with a prescribed total force on the spheres.
We evaluate the layer potential operators to an accuracy of $10^{-11}$ and the GMRES solve to a relative accuracy of $10^{-8}$.
We report the layer-potential operator setup time $T_{\text{setup}}$ and evaluation time $T_{\text{eval}}$
for on-surface target points (collocation points) during the solve and for off-surface target points when evaluating the final solution
on uniform grids matching the number of surface collocation points (only the subset that lies in the exterior of the spheres).
We report the percentage increase in the time compared to the corresponding free-space problem
for different problem sizes, measured in degrees of freedom $N_{\text{dof}}$.
The solution is computed on a single compute node, utilizing only one CPU socket with 32 cores.
We use one MPI process and 32 OpenMP threads with each thread bound to one core.

\begin{table}[htbp]
\centering
\renewcommand{\arraystretch}{1.5}
\setlength{\tabcolsep}{6pt}
\resizebox{\linewidth}{!}{%
\begin{tabular}{ll ccc ccc ccc ccc}
\toprule
\multicolumn{2}{l}{$N_{\text{dof}}$} & \multicolumn{3}{c}{$1.44\times 10^5$} & \multicolumn{3}{c}{$2.88\times 10^5$} & \multicolumn{3}{c}{$5.76\times 10^5$} & \multicolumn{3}{c}{$1.15\times 10^6$} \\
\cmidrule(lr){3-5} \cmidrule(lr){6-8} \cmidrule(lr){9-11} \cmidrule(lr){12-14}
\multicolumn{2}{l}{} & $T_{\text{free}}$ & $T_{\text{peri}}$ & \% change & $T_{\text{free}}$ & $T_{\text{peri}}$ & \% change & $T_{\text{free}}$ & $T_{\text{peri}}$ & \% change & $T_{\text{free}}$ & $T_{\text{peri}}$ & \% change \\
\midrule
\multirow{2}{*}{On-surface} & $T_{\text{setup}}$ (s)  & 11.69 & 13.11 & 12 & 23.32 & 25.53 & 10 & 50.08 & 52.71 &  5 & 93.75 & 98.29 &  5 \\
& $T_{\text{eval}}$ (s)                               &  0.45 &  0.52 & 16 &  0.88 &  1.13 & 28 &  1.75 &  1.87 &  7 &  3.13 &  3.39 &  8 \\
\midrule\addlinespace[0.3em]
\multirow{2}{*}{Off-surface} & $T_{\text{setup}}$ (s) &  7.79 &  8.84 & 13 & 15.49 & 16.78 &  8 & 31.91 & 33.32 &  4 & 63.00 & 65.73 &  4 \\
& $T_{\text{eval}}$ (s)                               &  0.31 &  0.45 & 45 &  0.82 &  1.03 & 25 &  1.61 &  2.22 & 38 &  2.38 &  2.77 & 16 \\
\bottomrule
\end{tabular}%
}
\caption{
    Periodization overhead in the setup and evaluation of the layer-potential operators for different problem sizes for the experimental setup in \cref{ss:periodization_overhead}.
    The setup time $T_{\text{setup}}$ and the evaluation time $T_{\text{eval}}$ for free-space and triply-periodic layer potential operators are reported for on-surface and off-surface targets, and the percent change is shown.
    The results are for a single MPI process and 32 OpenMP threads on a single CPU socket.
    Percentage changes are rounded to the nearest integer.
    As the problem size (in total degrees-of-freedom $N_{\text{dof}}$) increases, the periodization overhead as a percentage of the total time decreases.
    This is because periodization involves additional interactions close to the faces of the cubic domain, and for larger problems, these are a small fraction of the total interactions.
}
\label{tab:periodization_time}
\end{table}

\subsection{Scaling}
\begin{figure}[htbp]
    \centering
    \includegraphics[width=\linewidth]{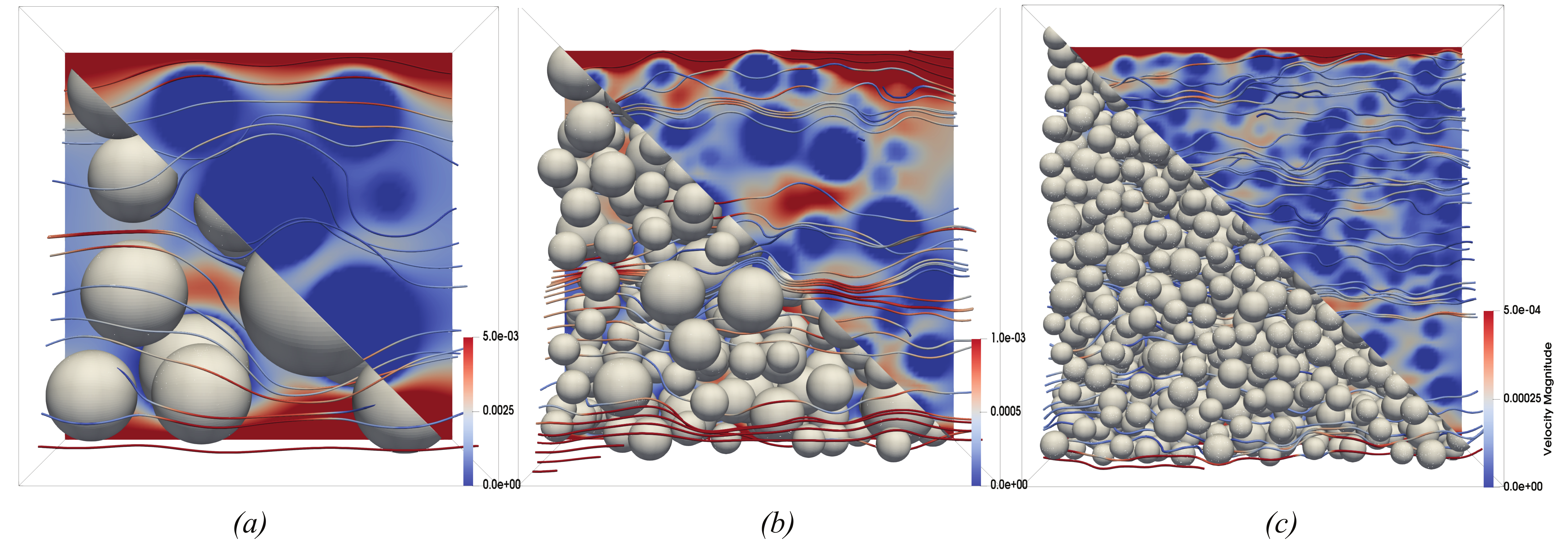}
    \caption{Disturbance-velocity streamlines in triply-periodic polydisperse sphere suspensions of 25 (left), 400 (center), and 2000 (right) spheres per unit cell, driven by a pressure gradient in the $+x$ direction. Each sphere is discretized with $N_p=4$ panels and $N_f=32$ Fourier modes; the BIE is solved to $10^{-9}$ tolerance. As the suspension density increases, the disturbance flow transitions from broad, well-separated wakes around individual spheres to a densely interconnected network of fine-scale, strongly interacting flow structures that fills the unit cell.}
    \label{fig:suspensions_and_flow}
\end{figure}

A primary motivation for this work is the simulation of dense, periodic polydisperse suspensions, where solvers must resolve a large number of degrees of freedom in the periodic cell while accurately capturing the flow in the narrow gaps between particles. As shown in \cref{fig:suspensions_and_flow}, the flow becomes increasingly complex as suspension density grows. In this section we demonstrate that the solver handles this regime with high accuracy at reasonable computational cost, with good scaling properties.
In \cref{tab:weak_strong_scaling}, we report the timing and select GMRES iteration counts corresponding to the total number of cores ($N_{\text{cores}}$) used. The degrees of freedom ($N_{\text{dof}}$) for each weak scaling example give the size of the unknown density $\bm{\mu}$ on all surface nodes in the unit cell.

We assess weak scaling using triply-periodic polydisperse suspensions, proportionally increasing the number of MPI processes ($N_{\text{proc}}$) with the number of particles while holding the volume fraction near 0.2 by reducing the maximum particle radius. For these systems we use $N_p=2$, $N_f=32$ and a quadrature tolerance of $10^{-14}$. \cref{tab:weak_strong_scaling}-left reports the setup time $T_{\text{setup}}$, which includes building the interaction tree but excludes operator precomputation, and the per-iteration GMRES evaluation time $T_{\text{eval}}$, averaged over multiple solves. 
For strong scaling, we fix the 2000-particle system at the highest discretization that fits in the 1\,TB memory on a single compute node ($N_p=3$, $N_f=48$, quadrature tolerance $10^{-12}$ per sphere) and increase the number of MPI processes. Results are given in \cref{tab:weak_strong_scaling}-right. 

\begin{table}[t]
\renewcommand{\arraystretch}{1.2}
\begin{minipage}[b]{.52\textwidth}
    \centering
    \begin{tabular}{rrrrr}
    \toprule
    $N_{\text{cores}}$ & $N_{\text{dof}}$ & $T_{\text{setup}}$ (s) & $T_{\text{eval}}$ (s) & \shortstack[r]{GMRES \\ iterations}\\
    \midrule
    32  & $4.80\times10^{4}$ & 11.08 & 2.13 & 101\\
    64  & $9.60\times10^{4}$ & 12.57 & 2.37 & 126\\
    128 & $1.92\times10^{5}$ & 13.57 & 2.40 & 128\\
    256 & $3.84\times10^{5}$ & 13.77 & 3.07 & 135\\
    512 & $7.68\times10^{5}$ & 15.10 & 3.95 & 142\\
    1024 & $1.54\times10^{6}$ & 14.00 & 3.31 & 141\\
    1280 & $1.92\times10^{6}$ & 16.31 & 3.51 & 148\\
    1600 & $2.40\times10^{6}$ & 18.56 & 3.73 & 161\\
    2048 & $3.07\times10^{6}$ & 16.34 & 3.94 & 162\\
    2240 & $3.36\times10^{6}$ & 16.33 & 4.09 & 160\\
    2432 & $3.65\times10^{6}$ & 15.85 & 4.19 & 160\\
    2560 & $3.84\times10^{6}$ & 17.21 & 4.37 & 165\\
    \bottomrule
    \end{tabular}
\end{minipage}\hfill
\begin{minipage}[b]{.44\textwidth}
    \centering
    \begin{tabular}{rrr}
    \toprule
    $N_{\text{cores}}$ & $T_{\text{setup}}$ (s) & $T_{\text{eval}}$ (s)\\
    \midrule
     32   & 473.45 & 26.59 \\
     64   & 230.02 & 12.31 \\
     128  & 124.68 &  6.77 \\
     256  &  63.42 &  4.20 \\
     512  &  32.28 &  2.14 \\
     1024 &  16.71 &  1.50 \\
     1280 &  13.69 &  1.41 \\
     1600 &  11.86 &  1.43 \\
     1920 &   9.80 &  1.27 \\
    \bottomrule
    \end{tabular}
\end{minipage}
\caption{Scaling results for triply-periodic polydisperse sphere suspensions. Time per GMRES iteration ($T_{\text{eval}}$) is averaged over multiple solves, the number of GMRES iterations is also listed. The total degrees of freedom ($N_{\text{dof}}$) in each weak scaling example grow proportionally with $N_{\text{cores}}$. Weak scaling (left) starts with a system of 25 spheres on 1 MPI process (32 cores), and increases the number of spheres and MPI processes proportionally while fixing the discretization parameters $N_p = 2$, $N_f = 32$ per sphere, quadrature tolerance $10^{-14}$, GMRES solved to $10^{-9}$ accuracy. Strong scaling (right) fixes the 2000-sphere system with $N_p=3$, $N_f =48$ per sphere, quadrature tolerance $10^{-12}$, GMRES solved to $10^{-8}$, and increases the number of MPI processes.
}
\label{tab:weak_strong_scaling}
\end{table}

Scaling efficiency is defined for each time metric $T \in \{T_{\text{setup}}, T_{\text{eval}}\}$ as
$$E_{\text{weak}}(N_{\text{proc}}) = \frac{T(1)}{T(N_{\text{proc}})},\qquad E_{\text{strong}}(N_{\text{proc}}) = \frac{T(1)}{N_{\text{proc}}\,T(N_{\text{proc}})},$$
with ideal efficiency equal to 1. In practice, inter-process communication and synchronization cause efficiency to decay with $N_{\text{proc}}$.

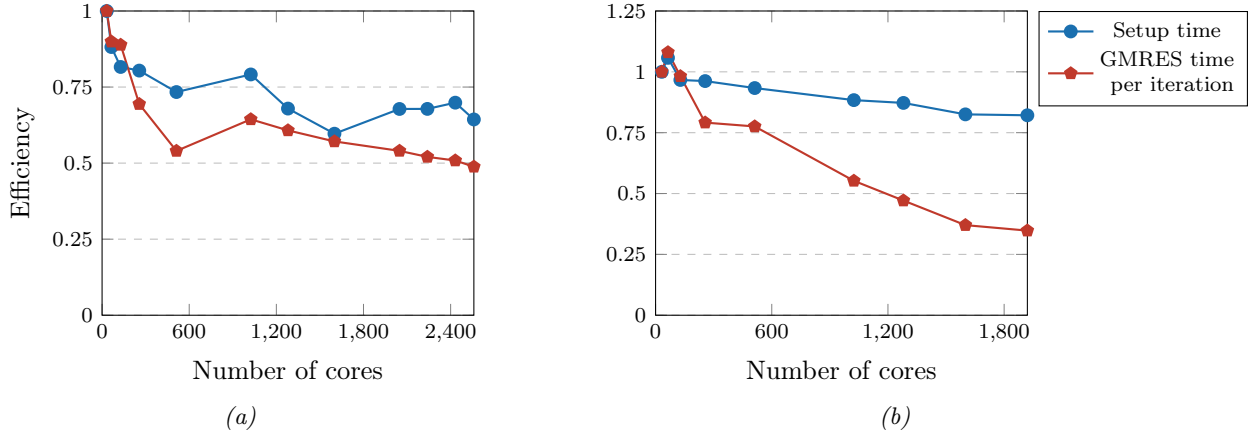
\begin{figure}[htbp]
\begin{subfigure}[b]{.4\textwidth}
    \centering
\begin{tikzpicture}
\begin{axis}[
    xlabel={Number of cores},
    ylabel={Efficiency},
    xmin=0, xmax=2560,
    ymin=0, ymax=1,
    xtick={0,600,1200, 1800, 2400, 3000},
    ytick={0.00,0.25,0.50,0.75,1.00},
    ymajorgrids=true,
    grid style=dashed,
    width=6.5cm,
    ticklabel style={font=\footnotesize},
]

\addplot[color={rgb,255:red,26;green,111;blue,175}, mark=*, mark size=2.2pt, line width=0.8pt]
    coordinates {
    (32,1)
    (64,0.881528997)
    (128,0.816020569)
    (256,0.804119146)
    (512,0.7335867701)
    (1024,0.7914159962)
    (1280,0.6792690032)
    (1600,0.596678446)
    (2048,0.6779801071)
    (2240,0.67820012)
    (2432,0.6986741182)
    (2560,0.6435223648)
    };
\addplot[color={rgb,255:red,192;green,57;blue,43}, mark=pentagon*, mark size=2.2pt, line width=0.8pt]
    coordinates {
    (32,1)
    (64,0.8992930702)
    (128,0.8887033143)
    (256,0.6940543594)
    (512,0.5400457718)
    (1024,0.64392546)
    (1280,0.6075838453)
    (1600,0.5712957631)
    (2048,0.5402581675)
    (2240,0.520466811)
    (2432,0.5085399674)
    (2560,0.4880713155)
    };

\end{axis}
\end{tikzpicture}
\caption{}
\label{fig:weak_scaling}
\end{subfigure}\hfill
\begin{subfigure}[b]{.5\textwidth}
\begin{tikzpicture}
\begin{axis}[
    xlabel={Number of cores},
    xmin=0, xmax=1920,
    ymin=0, ymax=1.25,
    xtick={0,600,1200, 1800},
    ytick={0.00,0.25,0.50,0.75,1.00,1.25},
    legend pos=outer north east,
    ymajorgrids=true,
    grid style=dashed,
    width=6.5cm,
    ticklabel style={font=\footnotesize},
    legend style={cells={align=right},font=\footnotesize}
]

\addplot[color={rgb,255:red,26;green,111;blue,175}, mark=*, mark size=2.2pt, line width=0.8pt]
    coordinates {
    (32,1)
    (64,1.057515812)
    (128,0.9666226671)
    (256,0.9620645842)
    (512,0.9333983115)
    (1024,0.8837849319)
    (1280,0.8722884384)
    (1600,0.8252691398)
    (1920,0.8214781787)
    };
    \addlegendentry{Setup time}

\addplot[color={rgb,255:red,192;green,57;blue,43}, mark=pentagon*, mark size=2.2pt, line width=0.8pt]
    coordinates {
    (32,1)
    (64,1.080242897)
    (128,0.9819605516)
    (256,0.7914634536)
    (512,0.7756533495)
    (1024,0.5522480278)
    (1280,0.4711603872)
    (1600,0.3702162542)
    (1920,0.3476437589)
    };
    \addlegendentry{GMRES time \\ per iteration}

\end{axis}
\end{tikzpicture}
\caption{}
\label{fig:strong_scaling}
\end{subfigure}
    \caption{Weak scaling efficiency (a) and strong scaling efficiency (b) for setup time (blue circles) and per-iteration GMRES evaluation time (red pentagons), normalized by the single-process baseline. }
    \label{fig:weak_strong_scaling}
\end{figure}

The solver sustains approximately 50\% weak scaling efficiency for both setup and evaluation across the tested range (\cref{fig:weak_scaling}), allowing larger systems to be computed with additional resources at modest overhead. Strong scaling efficiency (\cref{fig:strong_scaling}) exceeds 1 at small core counts because the single-socket baseline (32 cores, 1 MPI process) is memory-bandwidth-limited to one NUMA domain; adding a second MPI process on the other socket doubles the available bandwidth, producing apparent super-linear speedup. Efficiency then declines as the per-process workload shrinks and MPI communication overhead grows.
Efficiency declines to roughly 35\% at 1920 cores for the evaluation step, primarily because each process handles too small a share of the work for the MPI communication overhead to be amortized. To minimize wall-time fluctuations, all timings are averaged over multiple runs, each comprising several GMRES solves.

\section{Conclusions}
\label{sec:conclusions}
We have presented a scalable BIE framework for three-dimensional particulate Stokes flow in singly-, doubly-, and triply-periodic geometries. The central design principle is to work exclusively with the free-space Green's function while capturing far-image contributions through a hierarchical proxy sum built from the KIFMM equivalent surfaces. Convergence of this sum is guaranteed by the net-force-zero compatibility condition, which is enforced implicitly through the mean-density subtraction in the integral representation. The resulting periodization operator depends only on the periodic-box geometry and is independent of the kernel and of the surfaces inside the box; it is precomputed once and reused verbatim across the Stokeslet, stresslet, and rotlet without any additional per-kernel derivation.

Two structural limitations of prior free-space-based periodization schemes are avoided by design. The extended linear system (ELS) of earlier 2D schemes~\cite{marple2016fast,barnett2018unified}, in which proxy strengths are co-solved with wall densities, does not appear here: the periodization precomputation is decoupled from the embedded surfaces and is not reassembled when particle configurations change. The multi-layer near field required by the M2L-based scheme of~\cite{yan2018} is replaced by a single layer of image boxes, with all remaining images handled by the hierarchical proxy sum.

Numerical experiments validate spectral convergence in the number of Fourier modes and high-order convergence in the number of surface panels. The periodization parameters converge exponentially in both the multipole order $m$ and the number of hierarchy levels $N_{\text{levels}}$; in practice $m = 16$ and $N_{\text{levels}} = 45$ suffice for accuracies near machine precision. The solver handles dense polydisperse suspensions of up to 2000 particles with millions of degrees of freedom on distributed-memory architectures, sustaining approximately $50\%$ weak-scaling efficiency through 80 MPI processes.

The kernel-independent precomputation should extend without modification to other linear elliptic PDEs for which a free-space FMM is available, including the Laplace and linear elasticity equations and other kernels admitting kernel-independent FMMs, suggesting a reusable periodization infrastructure applicable across a broad class of integral-equation solvers.

\section{Acknowledgements} 
The authors thank Alex Barnett and Saibal De for helpful discussions. TL and SV acknowledge the support of NSF under grant DMS-2513346.
\bibliographystyle{plainnat}
\bibliography{references}
\end{document}